\documentstyle[preprint,aps,epsf]{revtex}
\begin{document}
\draft

\def\Oa{$O(a)$\ }

\title{
\vspace{-3.0cm}
\begin{flushright}  
{\normalsize UTHEP-404}\\
\vspace{-0.3cm}
{\normalsize UTCCP-P-66}\\
\vspace{-0.3cm}
{\normalsize June 1999 }\\
\end{flushright}
%
Two dimensional lattice Gross--Neveu model \\
with domain-wall fermions
}

\author{Taku Izubuchi $^{a}$\footnote{izubuchi@het.ph.tsukuba.ac.jp} 
and Kei-ichi Nagai $^{a,b}$\footnote{nagai@rccp.tsukuba.ac.jp}}
\address{$^a$ Institute of Physics,  University of Tsukuba,
         Tsukuba, Ibaraki 305-8571, Japan}
\address{$^b$ Center for Computational Physics,  University of Tsukuba,
         Tsukuba, Ibaraki 305-8577, Japan}

\date{\today}
\maketitle

\begin{abstract}
We investigate 
the two dimensional lattice Gross--Neveu model
in large flavor number limit
using the domain-wall fermion formulation,
as a toy model of lattice QCD.
We study  nonperturbative behavior
of the restoration of chiral symmetry 
of the domain-wall fermions
as the extent of the extra dimension $(N_s)$ is increased
to infinity.
We find the the parity broken phase (Aoki phase) for finite $N_s$, 
and study the phase diagram, which is related to 
the mechanism of the chiral restoration in $N_s\to\infty$ limit.
The continuum limit is taken and $O(a)$ scaling violation of observables
vanishes in $N_s\to\infty$ limit.
We also examine the systematic dependencies of observables to 
the parameters. 
\end{abstract}
\narrowtext
\clearpage
\section{Introduction}
\label{sec:intro}
Chiral symmetry is one of the important properties
to understand the hadron physics 
and the phase transition in the thermodynamics of field theories.
Pions are regarded as the  pseudo Nambu-Goldstone bosons 
associated with the spontaneous breakdown of chiral symmetry.
Physics of the phase transition between the confining phase (Hadron phase)
and the deconfining phase (Quark-Gluon Plasma phase) in QCD is 
interested from the theoretical and experimental point of view.
The lattice field theory is one of the most powerful tool 
for such important physics beyond perturbations.

However there is a problem to define chiral symmetry 
on a lattice.
The definition of chiral symmetry on the lattice 
is one of long-standing problems 
of the lattice field theory.
This problem is called as ``no-go
theorem'' \cite{nogo1,nogo2}:
the unwanted species of fermions appear 
in chiral symmetric theory on the lattice.
To avoid this,
chiral symmetry has to be broken 
by adding the Wilson term to the Lagrangian\cite{wilson,wilson1}.
This formulation is known as Wilson fermion (WF).
In order to obtain the chiral symmetric theory in the continuum limit,
one has to fine tune the quark mass parameter to cancel the additive 
quantum correction, which is a non trivial task in numerical simulations.
Besides this fine tuning difficulty,
the physical prediction from WF has $O(a)$ scaling violation
due to the absence of chiral symmetry,
where $a$ is the lattice spacing.
One must calculate on a fine lattice to get precise predictions
in WF.

Few years ago,
the domain-wall fermion (DWF) is proposed \cite{domain1,domain2}
as the new formulation of lattice fermion.
DWF is formulated as $D+1$ dimensional 
Wilson fermion 
with the free boundary condition 
in the extra dimension \cite{domain2}.
When the extra dimension becomes infinite,
DWF has a massless fermion with chiral symmetry.
Because of this property,
the current quark mass term
in DWF receives  multiplicative renormalization
in contrast with WF,
and this fermion formulation does not need the fine tuning
in order to restore chiral symmetry 
on the lattice \cite{pert1,pert2}.
The $O(a)$ scaling violation is expected to vanish in large $N_s$ limit,
which means smaller scaling violations than in WF.
Thus DWF has desirable property for defining 
the lattice fermion especially 
for chiral symmetric or near massless cases.

The numerical simulations of the domain-wall QCD
(DWQCD) have been already carried out \cite{Simulations1,Simulations2}.
As a trade off of the above ideal properties,
one needs a large amounts of CPU time for computer simulations of DWQCD.
Furthermore the new model has larger number of parameters 
than that of WF, and  results of simulations have complicated 
dependencies to parameters.

To clarify the nonperturbative properties of DWF, 
we examine the formalism for the solvable theory, 
Gross-Neveu model (GN model) in two dimensions, 
which shares common natures with QCD. 
Our main purpose in this paper is 
to understand the restoration of chiral symmetry.
For finite $N_s$, for which the lattice simulations are performed, 
we will see that chiral symmetry is broken,
and the model can be seen as ``improved Wilson fermion".
DWF for finite $N_s$ has the parity broken phase 
similar to WF\cite{aoki,umemura,kaneda}. 
We also study the continuum limit of DWF model.  
We will see the $O(a)$ becomes small 
as $N_s$ is increased and vanishes in the limit $N_s\to\infty$.

This paper is organized as follows:
In Sec.\ref{sec:GNpot},
after a brief review of the GN model in the continuum,
the lattice GN model with DWF is formulated
and the effective potentials in large flavor ($N$) limit are calculate
when the extent of extra dimension $(N_s)$ is both infinite and finite.
In Sec.\ref{sec:effpnsinf}
we calculate the continuum limit 
in infinite and finite $N_s$ case 
and discuss the restoration of chiral symmetry.
We will see that the model has  chiral symmetry  on the lattice for 
infinite $N_s$, while the symmetry is restored only in the 
continuum limit with fine tuning for finite $N_s$.
In Sec.\ref{sec:Aokiphase}
we analyze the structure of the chiral phase boundary
between the parity symmetric and the parity broken phase (Aoki phase)
for the finite lattice spacing by solving the gap equation
and discuss the necessity of fine tuning
to restore the chiral symmetry.
In Sec.\ref{sec:paradep} 
we study the parameter dependences 
of the lattice observables and in Sec.\ref{sec:Aerr} 
we discuss the way of taking the continuum limit.
We will see that the correct  continuum limit are taken from 
the lattice model and $O(a)$ scaling violation 
vanishes for $N_s\to\infty$ limit.
We conclude with a summary in Sec. \ref{sec:conclusion}.

\section{Action and The Effective Potential}
\label{sec:GNpot}

\subsection{Continuum Gross--Neveu model}
\label{ssec:Continuum}
The investigation of the GN model is a good test 
\cite{aoki,gnlat,VTK,IchNag,FiniteMuGN} 
for the nonperturbative behavior of QCD 
since the two theories share common properties: 
the feature of asymptotic freedom, 
chiral symmetry and its spontaneous breakdown.

The two dimensional continuum Gross-Neveu model 
in Euclidean space is defined by the action
\begin{equation}
S = \int dx^2 \{
\bar\psi(\gamma_\mu\partial_\mu + m)\psi
-{g^2\over 2N}[ (\bar\psi\psi)^2 + (\bar\psi i\gamma_5\psi)^2 ]
\}~, 
\label{eq:contaction}
\end{equation}
where $\psi$ is an $N$-component fermion field.
The effective potential in large $N$ limit in the continuum 
theory is given as 
\begin{equation}
V_{eff} = - m \sigma + \frac{1}{4 \pi} 
\left( \sigma^2 + \Pi^2 \right) 
\ln \frac{\sigma^2 + \Pi^2}{e \Lambda^2} \, ,
\label{eq:contVeff}
\end{equation}
where $m$ is the renormalized mass and $\Lambda$ is the scale parameter.
If the momentum integration is regularized by a cutoff, $M$,
the renormalization for bare mass $m_0$ and 
the bare coupling constant $g^2$ are 
\begin{eqnarray}
m &=& {m_0 \over g^2} ~,\\
{1\over g^2} &=& {1\over 2\pi} \ln {M^2\over\Lambda^2} ~,
\end{eqnarray}
where the latter shows the asymptotically free. 
(We set the renormalized coupling constant to unity)

The auxiliary field $\sigma$ and $\Pi$ relate to the fermion condensations
by equation of motion
\begin{equation}
\sigma=m_0 -\frac{g^2}{N} \bar\psi\psi,~~~
\Pi=\frac{g^2}{N} \bar\psi i\gamma_5\psi~.
\end{equation}
When $m_0$ vanishes the model shows chiral symmetry, 
which is expressed by the $O(2)$ rotational invariance of 
effective action (\ref{eq:contVeff}) in $(\sigma,\Pi)$ space.
The stationary point of the effective action (\ref{eq:contVeff})
satisfies $\sigma^2+\Pi^2=\Lambda^2$, which manifests 
the spontaneous breakdown of chiral symmetry.

\subsection{Lattice model with DWF}
\label{sebsec:setup}

DWF is formulated as Wilson fermion (WF)
in $D+1$ dimensions, or equivalently $N_s$ flavor
WF with a flavor mixing,
which has negative Wilson term
obeying the free boundary condition 
on the edges in the extra dimension\cite{domain2}.

The action of DWF is given as follows:
\begin{equation}
S_{\rm free} = a^2 \sum_{s, t} \sum_{m, n} \overline{\psi}(m,s) 
D^{free}(m, s ;  n , t) \psi(n,t) ,
\label{eq:action}
\end{equation}
where 
\begin{eqnarray}
&& D^{free}(m , s ; n , t) = \sum_{\mu = 1}^{2} 
\sigma_{\mu} C_{\mu}(m,n) \delta_{s,t} - W(m,n)
\delta_{s,t}
+ P_R \delta_{s+1,t} \delta_{m,n} 
+ P_L \delta_{s,t+1} \delta_{m,n} ,\\
&& C_{\mu}(m,n) = \frac{a_s}{2 a} 
\left[ \delta_{m+\hat{\mu},n} - \delta_{m-\hat{\mu},n}
\right] , \\
&& W(m,n) = (1 - M) \delta_{m,n} + \frac{r a_s}{2 a} \left[ 
 2 \delta_{m,n} - 
\delta_{m+\hat{\mu},n} - \delta_{m,n+\hat{\mu}}  \right]
 \, .
\label{eq:free}
\end{eqnarray}

$P_{R/L}=(1 \pm \gamma_5)/2$ are the projection operators
into the right- and the left-handed mode,
$a$ and $a_s$ are the lattice spacing in two dimensions 
and the third dimensions respectively.
$\sigma_\mu$'s are defined as $\sigma_1=i$ and $\sigma_2=1$
in two dimensions,
$s$ and $t$ are the indecies of extra dimension 
with $1 \leq s, t \leq N_s$.
Here, $r ( >0 )$ is Wilson coupling constant
and $M$ is the domain-wall mass height (DW-mass).
The boundary condition in the third direction
takes the free boundary condition.
In the followings,
we take $a=a_s$ and $r=1$ for simplicity.

If one sets $0 < M < 2$, 
there is {\it single} light Dirac fermion
in the spectrum of this free action whose right(left)-handed 
components stays near $s=1(N_s)$,
\begin{equation}
q(n)= P_R \psi(n,s=1)+P_L\psi(n,s=N_s)~.
\end{equation}
The mass of this light quark, $q(n)$, 
is exponentially suppressed for large $N_s$,
\begin{equation}
m_q a  \sim (1-M)^{N_s} ~.
\end{equation}
Wilson term in the action avoids the species doubling problem.
The doublers and other $N_s-1$ bulk fermions acquire 
the cut-off order mass and 
decouple from low energy physics.

As a toy model of the lattice DWQCD, 
we define the two dimensional lattice GN model with $N$ flavors:
\begin{equation}
S = S_{free} 
+ a^2\sum_n a m_f \bar{q}(n)q(n)
- a^2 \sum_n 
\left[\frac{g_{\sigma}^2}{2 N}
\left\{\bar{q}(n)q(n)\right\}^2
+\frac{g_{\pi}^2}{2 N}
\left\{\bar{q}(n)i\gamma_5 q(n)\right\}^2\right] ~,
\label{eq:GNaction}
\end{equation}
where we abbreviate
$\overline{\psi}(n,s)\psi(n,t) = \sum_{i=1}^N
\overline{\psi}^i(n,s) \psi^i(n,t)$.
$m_f$ is the ``current quark mass'' in order to give 
the mass to the fermion\footnote{
The parameters, $M$, $r$ and $m_f$, in this paper 
are opposite sign with the Ref.\cite{Vra}.}.
In perturbation of DWQCD, 
$m_f$ receives the multiplicative renormalization 
in $N_s\to\infty$ case\cite{pert1,pert2}.
DW-mass $M$, on the other hand,
receives the additive renormalization\cite{pert1,pert2},
because $M$ corresponds to Wilson mass term.
We will see that two different couplings, 
$g^2_\sigma$ and $g^2_\pi$, are needed 
for chiral symmetry in general.

The action (\ref{eq:GNaction}) can be rewritten 
into the following equivalent action 
using the auxiliary fields $\sigma(n)$ and $\Pi(n)$:
\begin{eqnarray}
S &=& S_{\rm free} + a^2 \sum_{n}\bar{q}(n) 
\left\{ a \sigma(n) + i \gamma_5 a \Pi(n) \right\} q(n)
\nonumber \\
&+& \quad a^2 \sum_{n} \left[ \frac{N}{2 g_{\sigma}^2} 
\left\{\sigma(n) - m_f \right\}^2 + \frac{N}{2 g_{\pi}^2} \Pi(n)^2 
\right]~.
\label{eq:GNterm}
\end{eqnarray}
The auxiliary fields are related to condensations of fermion;
\begin{eqnarray}
 \sigma(n) =  m_f - \frac{g_{\sigma}^2}{N} \bar{q}(n)
q(n) \quad , \quad
 \Pi(n) =  - \frac{g_{\pi}^2}{N} \bar{q}(n) i \gamma_5
q(n) \, .\label{eq:pi}
\end{eqnarray}
from the equations of motion.

\subsection{the effective potential} 
\label{subsec:caleffpot}
We calculate the effective potential 
of the two dimensional lattice GN model
in large $N$ limit.
In large $N$ limit, 
in which the quantum fluctuation of 
$\sigma(n)$ and $\Pi(n)$ is suppressed and 
the mean field approximation,
$\sigma(n) \to \sigma$ and $\Pi(n) \to \Pi$, 
becomes exact.
the effective potential of GN model
is obtained by exponentiating the fermion determinant
which is calculated by the integration 
of the fermion fields $\psi$:
\begin{eqnarray}
Z &=& \int [d \psi ] [ d \overline{\psi}]
e^{- a^2 \sum \overline{\psi} D \psi}  
e^{- a^2 \sum [ \frac{N}{2 g_{\sigma}^2} 
(\sigma - m_f)^2 + \frac{N}{2 g_{\pi}^2} \Pi^2 ]} 
\nonumber \\
&=&  \det D(\sigma,\Pi) \, e^{- a^2 \sum [ \frac{N}{2 g_{\sigma}^2} 
(\sigma - m_f)^2 + \frac{N}{2 g_{\pi}^2} \Pi^2 ]}
= e^{- a^2 V_{eff}} \, .
\label{eq:veff}
\end{eqnarray}
This determinant can be calculated by employing
the technique of the propagator matrix \cite{truncate1,truncate2}.
The action introduced in the previous subsection is rewritten 
by the matrix representation,
in the momentum space.
From (\ref{eq:GNterm}),
\begin{eqnarray}
&&S = \bar\psi D(\sigma,\Pi)\psi = a^2 \sum_{s,t} \sum_{m,n} 
\left( \psi^{\dag}_L(m,s) , \psi^{\dag}_R(m,s) \right)
\left[
\left(
\begin{array}{cc}
C^{\dag} & -W \\
-W & - C
\end{array}
\right) \delta_{s,t} + 
\left(
\begin{array}{cc}
0 & 0 \\
1 & 0
\end{array}
\right) \delta_{t,s+1} \right. \nonumber \\
&&\left. + \left(
\begin{array}{cc}
0 & 1\\
0 & 0
\end{array}
\right) \delta_{s,t+1} + 
\left(
\begin{array}{cc}
0 & 0 \\
a \omega & 0
\end{array}
\right) \delta_{s,N} \delta_{t,1} + 
\left(
\begin{array}{cc}
0 & a \omega^{\dag} \\
0 & 0
\end{array}
\right) \delta_{s,1} \delta_{t,N_s}
\right]
\left(
\begin{array}{c}
\psi_L(n,t) \\
\psi_R(n,t)
\end{array}
\right) \, ,
\label{eq:action2}
\end{eqnarray}
where 
$C^{\dag} =   \sin(p_1 a) + i \sin(p_2 a) $, 
$\omega = \sigma + i \Pi$
and
$W = (1-M) +  \sum_{\mu=1}^2 [1 - \cos (p_\mu a)] $.
After some calculations similar to Ref.\cite{truncate1,truncate2},
we have
\begin{equation}
\det D(\sigma,\Pi) 
=
\det \left[
\left(
\begin{array}{cc}
a  \omega & 0 \\
0 & 1
\end{array}
\right) 
- T^{-N_s} 
\left(
\begin{array}{cc}
1 & 0 \\
0 &  a \omega^{\dag}
\end{array}
\right) \right] \, ,
\label{eq:det}
\end{equation}
where $T$ is the transfer matrix along the extra dimension
defined by
\begin{equation}
T = e^{a_s H_s} = \left(
\begin{array}{ccc}
\frac{\displaystyle{1}}{\displaystyle{W}} &,& -\frac{\displaystyle{1}}{\displaystyle{W}} C \\
-C^{\dag} \frac{\displaystyle{1}}{\displaystyle{W}} &,& W + C^{\dag} 
\frac{\displaystyle{1}}{\displaystyle{W}} C
\end{array}
\right) \quad , \quad 
T^{-1}  = \left(
\begin{array}{ccc}
C^{\dag} \frac{\displaystyle{1}}{\displaystyle{W}} C + W
&,& C 
\frac{\displaystyle{1}}{\displaystyle{W}} \\
 \frac{\displaystyle{1}}{\displaystyle{W}} C^{\dag} &,&  
\frac{\displaystyle{1}}{\displaystyle{W}}
\end{array}
\right) \, .
\end{equation}
$H_s$ is the Hamiltonian along to the extra dimension.
The overall factor in Eq. (\ref{eq:det}) is omitted.
Diagonalized $T^{-1}$, whose eigenvalues are $\lambda$ and $1/\lambda$,
we obtain a explicit formula for the determinant,
\begin{equation}
\det D (\sigma , \Pi) = 
\prod_{p_\mu}
\left[  F(M,N_s,p_\mu) a^2 (\sigma^2 + \Pi^2)  
+ G(M,N_s,p_\mu) a \sigma 
+ H(M,N_s,p_\mu) 
\right] \, ,
\label{eq:det2}
\end{equation}
where
\begin{eqnarray}
&& H(M,N_s,p_\mu) = \frac{-1}{2f} 
\left[ -(\lambda^{N_s} - \lambda^{-N_s}) (2 - h) 
+ (\lambda^{N_s} + \lambda^{-N_s}) f \right] \, , 
\label{eq:funcH}\\
&& F(M,N_s,p_\mu) = \frac{-1}{2f}
\left[ (\lambda^{N_s} - \lambda^{-N_s}) (2 - h) 
+ (\lambda^{N_s} + \lambda^{-N_s}) f \right] \, , 
\label{eq:funcF}\\
&& G(M,N_s,p_\mu) =  \frac{-1}{2f} \left[ - 4 f \right] = 2\, , 
\label{eq:funcG}\\
&& h = 1 + W^2 + \bar{p}^2 \quad , \quad f = \sqrt{-4 W^2 + h^2} , \\
&& \lambda \,
= \frac{1}{2 W} \left[ h + f \right]\,\,  
( \left| \lambda\right| > 1) \quad  , \quad
\frac{1}{\lambda} 
= \frac{1}{2 W} \left[ h - f \right]\,\,  
(\left|\lambda \right|< 1) \, , \\
&& W = (1-M) + \sum_{\mu=1}^2
\left[1 - \cos(p_\mu a) \right] \quad , \quad 
\bar{p}^2 = \sin^2(p_1 a) + \sin^2(p_2 a)  \, .
\end{eqnarray}
Substituting (\ref{eq:det2}) into (\ref{eq:veff}),
we obtain the effective potential of the two dimensional lattice GN model 
in the large $N$ limit: 
\begin{eqnarray}
&& V_{eff} = \frac{1}{2 g_{\sigma}^2} \left( \sigma - m_f \right)^2
 +  \frac{1}{2 g_{\pi}^2} \Pi^2 - I(\sigma,\Pi,M,N_s)  
\label{eq:pot} \\
&& I (\sigma,\Pi,M,N_s) = \int_{-\pi/a}^{\pi/a} \frac{d^2 p}{(2 \pi)^2}
\ln \left[ F a^2 (\sigma^2 + \Pi^2) + G a \sigma + H
\right] .
\label{eq:effpot}
\end{eqnarray}
This effective potential is symmetric 
under $M \rightarrow 6-M$. 

 Here we comment about the absence of the 
bosonic fields (Pauli-Villars fields) in our model
which is employed in (full) QCD\cite{pert1,Simulations1,Simulations2}.
Since the gauge field in QCD equally interacts with 
not only light quark field, $q(n)$,
but also heavy (bulk) fermions $\psi(n,s)$,
one should introduce the Pauli-Villars boson to subtract this 
heavy fermion effects.
In GN model $\sigma$ and $\Pi$ play analogous role to 
the gauge field in full QCD except that they couples with $q(n)$ only.
Thus it is not necessary to introduce the subtraction in our model.
One the other hand one could think about a new four Fermi interaction model, 
in which whole $\psi(n,s),s=1,\cdots,N_s$ equally couple to auxiliary 
fields, $\sigma$ and $\Pi$. However these fields are constituted from 
both light and heavy fermions and don't show the chiral property 
in such a new model.

\section{Chiral symmetry restoration}
\label{sec:effpnsinf}

First we show how chiral symmetry restores in this model in the case of
{\it both} $N_s=\infty$ and $N_s < \infty$ by examining 
the effective potential given in previous section.

We will see that for infinite $N_s$ case chiral symmetry is exact
even for finite lattice spacing, $a>0$, 
without fine tuning for bare mass parameters. 
The situation for finite $N_s$ case, on the other hand, 
is much like that of Wilson fermion action.
The continuum limit has to be taken, and at same time, 
the bare mass parameter must be tuned finely for 
chiral symmetric effective potential for $N_s<\infty$.

\subsection{The effective potential in ``$N_s = \infty$'' case}
\label{subsec:Nsinf}
In this subsection we calculate the expression of the effective potential
for $N_s=\infty$ case.

For large $N_s$, one easily sees that
the dominant contributions for function $I(\sigma,\Pi)$ in the 
effective potential (\ref{eq:effpot}) are the functions 
``$F$'' and ``$H$'',
which behave as $\lambda^{N_s}$ in $N_s \rightarrow \infty$ limit,
from Eqs. (\ref{eq:funcH}) and (\ref{eq:funcF}).
The chiral breaking term ``$G a\sigma$'' in (\ref{eq:effpot})
can be ignore in the limit $N_s\to\infty$ 
and  (\ref{eq:effpot}) could be written as
\begin{equation}
I(\sigma,\Pi) = \int_{-\pi/a}^{\pi/a} \frac{d^2 p}{(2 \pi)^2}
\ln \left[ H + F a^2 (\sigma^2 + \Pi^2) 
\right].
\label{eq:INsInf}
\end{equation}
So the effective potential (\ref{eq:pot}) is a function of 
$\sigma^2+\Pi^2$, which is invariant under the $O(2)$ rotation
if $g_\sigma^2=g_\pi^2$.
We emphasize that this is chiral symmetry even before 
taking the continuum limit, $a\to 0$.

The continuum limit of the effective potential can be evaluated 
by separating  divergent parts and finite parts from (\ref{eq:INsInf}).
Rewriting (\ref{eq:INsInf}) into a integration form
\begin{equation}
I(\sigma,\Pi) = \int_{0}^{\sigma^2+\Pi^2} d \rho 
K(\rho) \quad , \quad 
K(\rho) = \int_{-\pi/a}^{\pi/a} \frac{d^2 p}{(2 \pi)^2}
\frac{1}{\frac{\displaystyle{H}}{\displaystyle{F a^2}} +
\rho} \, ,
\label{eq:kernel1}
\end{equation}
one can pick up the the divergent part in $a \to 0$, near
zero fermion momentum, $p_\mu a = (0 , 0)$, for $\rho a^2 \sim 0$.
Since the divergent part of $K(\rho)$ behaves as
\begin{equation}
\frac{H}{F a^2} \rightarrow \frac{1}{M^2 (2-M)^2}
\sum_{\mu=1}^{2} p_\mu^2 \,, 
\label{eq:HFNinf}
\end{equation}
in $a \to 0$ limit,
the function $K(\rho)$ becomes 
\begin{equation}
K(\rho) = \int_{-\pi/a}^{\pi/a} \frac{d^2 p}{(2 \pi)^2}
\frac{1}{f_M^{-2} \sum_{\mu=1}^{2} p_\mu^2 + \rho} 
+ C_0 (M,N_s) \, ,
\label{eq:kernel2}
\end{equation}
where $f_M = M (2-M)$ and
\begin{equation}
C_0(M,N_s) = \int_{-\pi}^{\pi} \frac{d^2 \xi}{(2 \pi)^2}
\frac{f_M^{-2} \sum_\mu \xi_\mu^2 -
\frac{\displaystyle{H}}{\displaystyle{F}}
}{\frac{\displaystyle{H}}{\displaystyle{F}} 
f_M^{-2} \sum_\mu \xi_\mu^2} \, ,
\label{eq:coef1}
\end{equation}
with $\xi_\mu = p_\mu a$.
The factor $f_M$ appears as the normalization factor 
of the propagator: 
$\langle q \bar q\rangle \sim f_M (i p_\mu\gamma_\mu)^{-1}$.
The wavefunction of the massless eigenmode has finite width
in the $s$ direction, and $f_M$ is the ratio of $q(n)$ to the 
the zero mode.

In $N_s=\infty$ case,
when $0<M<2$,
only the momentum around $p_\mu a =(0,0)$ dominates in (\ref{eq:INsInf})
and the contributions of doublers, 
$p_\mu a = (\pi,0),(0,\pi)$ and $(\pi,\pi)$,
are removed completely.
This means that
the doublers  decouple from the physical spectrum.
For $2<M<4$ the momenta $p_\mu a = (\pi,0)$ and $(0,\pi)$
become physical poles in the momentum integral,
while the remaining mode at $p_\mu a =(\pi,\pi)$ is dominant for $4<M<6$.
In these two regions of $M$,
the normalization factor $f_M$
becomes $(M-2)(4-M)$ and $(M-4)(6-M)$ respectively.
For $M<0$ and $M>6$,
the no physical pole emerges.

By evaluation of the first term in the right hand side
in Eq. (\ref{eq:kernel2}),
we find 
\begin{equation}
K(\rho) = \frac{f_M^2}{4 \pi} 
\ln \frac{1}{a^2 f_M^2 \rho} + \hat{C_0}(M,N_s),
\end{equation}
where the new constant $\hat{C_0}$ is defined by 
$\hat{C_0} = C_0 + C_0^{\prime}$ with
\begin{equation}
\int_{-\pi/a}^{\pi/a} \frac{d^2 p}{(2 \pi)^2}
\frac{1}{f_M^{-2} \sum_{\mu=1}^{2} p_\mu^2 + \rho} \equiv
\frac{f_M^2}{4 \pi} \ln \frac{1}{a^2 f_M^2 \rho} +
C_0^{\prime}(M,N_s) .
\end{equation}
By substituting this expression into Eq. (\ref{eq:kernel1}),
we obtain
\begin{equation}
I = - \frac{f_M^2}{4 \pi} \left(\sigma^2 + \Pi^2 \right)
\ln \frac{a^2 f_M^2 \left(\sigma^2 + \Pi^2 \right)}{e}
+ \hat{C_0} \left(\sigma^2 + \Pi^2 \right).
\end{equation}

Therefore 
the continuum limit of the effective potential for $N_s=\infty$ case is
\begin{equation}
V_{eff} =  - \frac{m_f }{g^2}\sigma
+ \left(\frac{1}{2 g^2} - \hat{C_0}  
+ \frac{f_M^2}{4 \pi} \ln a^2 \right) 
\left(\sigma^2 + \Pi^2 \right) 
+ \frac{f_M^2}{4 \pi} \left(\sigma^2 + \Pi^2 \right) 
\ln \frac{f_M^2 \left(\sigma^2 + \Pi^2 \right)}{e}.
\label{eq:potential1}
\end{equation}
We find that the two coupling constants can be same as  each other
$g^2=g_\sigma^2 = g_\pi^2$ from above result, 
which is in contrast to the GN model with WF.
As mentioned above,
since the effective potential has chiral symmetry for finite lattice spacing,
the fine tuning is unnecessary for chiral symmetric continuum limit.

After redefining for $\sigma$ and $\Pi$ 
such as $\sigma_R = f_M \sigma$ and $\Pi_R = f_M \Pi$,
we renormalize the coupling constant $g$  and $m_f$ as follows:
\begin{eqnarray}
&&\frac{1}{2 g^2} = \hat{C_0} + \frac{f_M^2}{4 \pi} \ln
\frac{1}{a^2 \Lambda^2} \, , 
\label{eq:coupling1} \\
&& m = \frac{m_f}{f_M g^2}
= m_f \left(\frac{2}{f_M} \hat{C_0} + \frac{f_M}{2 \pi} 
\ln\frac{1}{a^2 \Lambda^2} \right) \, , 
\label{eq:qmass}
\end{eqnarray}
where $\Lambda$ is the scale parameter 
and $m$ is the renormalized mass parameter.
One realizes that the current quark mass term, $m_f$, receives 
the multiplicative renormalization (\ref{eq:qmass}),
similar to the perturbation theory in infinitely large $N_s$.
With this choice of scaling relations
the continuum limit of the effective potential for $N_s=\infty$
is finally given as
\begin{equation}
V_{eff} = - m \sigma_R + \frac{1}{4 \pi} 
\left( \sigma_R^2 + \Pi_R^2 \right)
\ln \frac{\sigma_R^2 + \Pi_R^2}{e \Lambda^2} ,
\label{eq:potential2}
\end{equation}
which is identical with the continuum theory (\ref{eq:contVeff}).

\subsection{Chiral restoration in the continuum limit 
($N_s$ finite case)}
\label{subsec:Nsfincont}

As we have seen in previous subsection, 
the limit $N_s \to \infty$ guarantees
chiral symmetry on a lattice.
Nevertheless, to understand the behavior of finite $N_s$ theory 
is important for lattice QCD simulations using DWF.

Let us start the finite $N_s$ analysis  
with the effective potential (\ref{eq:pot}).
The chiral breaking term, $G a\sigma$, in (\ref{eq:effpot}) 
has the important role for finite $N_s$ 
and the GN model does not have chiral symmetry any more.
Chiral symmetry can be restored in the continuum limit
with fine tuning.
In this sense, finite $N_s$ model is similar to WF.
In order to restore the chiral symmetry 
without fine tuning,
one should take the limit $N_s\to\infty$ {\it before}
taking the continuum limit to obtain chiral symmetry on lattice.

The continuum limit for finite $N_s$ can be taken following the same 
procedures in the previous subsection.
The difference from the calculation
in $N_s\to\infty$ is a necessity to shift the auxiliary field: 
\begin{equation}
\sigma^{\prime} = 
\sigma - \frac{\displaystyle{(1-M)^{N_s}}}{\displaystyle{a}} ~.
\label{eq:sigmaShift}
\end{equation}
The effective potential becomes as follows:
\begin{eqnarray}
&& V_{eff} = \frac{1}{2 g_{\sigma}^2} 
\left\{ \sigma^{\prime} - \left( m_f -
\frac{(1-M)^{N_s}}{a} \right)
  \right\}^2
 +  \frac{1}{2 g_{\pi}^2} \Pi^2 - I(\sigma,\Pi)~,\\  
&& I(\sigma,\Pi) = \int_{-\pi/a}^{\pi/a} \frac{d^2 p}{(2 \pi)^2}
\ln \left[ H^{\prime} + F a^2 ({\sigma^{\prime}}^2 + \Pi^2) 
+ G^{\prime} a \sigma^{\prime} \right] ,
\label{eq:effpotR}
\end{eqnarray}
where
\begin{equation}
H^{\prime} = H + F (1-M)^{2 N_s} + G (1-M)^{N_s} \quad , \quad
G^{\prime} = G + 2 F (1-M)^{N_s} \, .
\end{equation}
Expanding $I(\sigma,\Pi)$ into a power series of lattice spacing:
\begin{eqnarray}
&& I = I_0 + I_1 + I_2 + \cdots \\
&&I_0 = \int_{-\pi/a}^{\pi/a} \frac{d^2 p}{(2 \pi)^2}
\ln \left[H^{\prime} + F a^2 ({\sigma^{\prime}}^2 + \Pi^2) \right] ,
\label{eq:I0R} \\
&&I_n = - \frac{(-1)^n}{n} 
\int_{-\pi/a}^{\pi/a} \frac{d^2 p}{(2 \pi)^2}
\left[ \frac{G^{\prime} a \sigma^{\prime}}{H^{\prime} 
+ F a^2 ({\sigma^{\prime}}^2 + \Pi^2)} \right]^n . \quad (n \geq 1)
\label{eq:InR}
\end{eqnarray}
One can see that $I_0$ preserve $O(2)$ rotation in $(\sigma',\Pi)$ plane,
while $I_n (n\geq 1)$ break it. 

\begin{figure}
\centerline{\epsfxsize=10cm \epsfbox{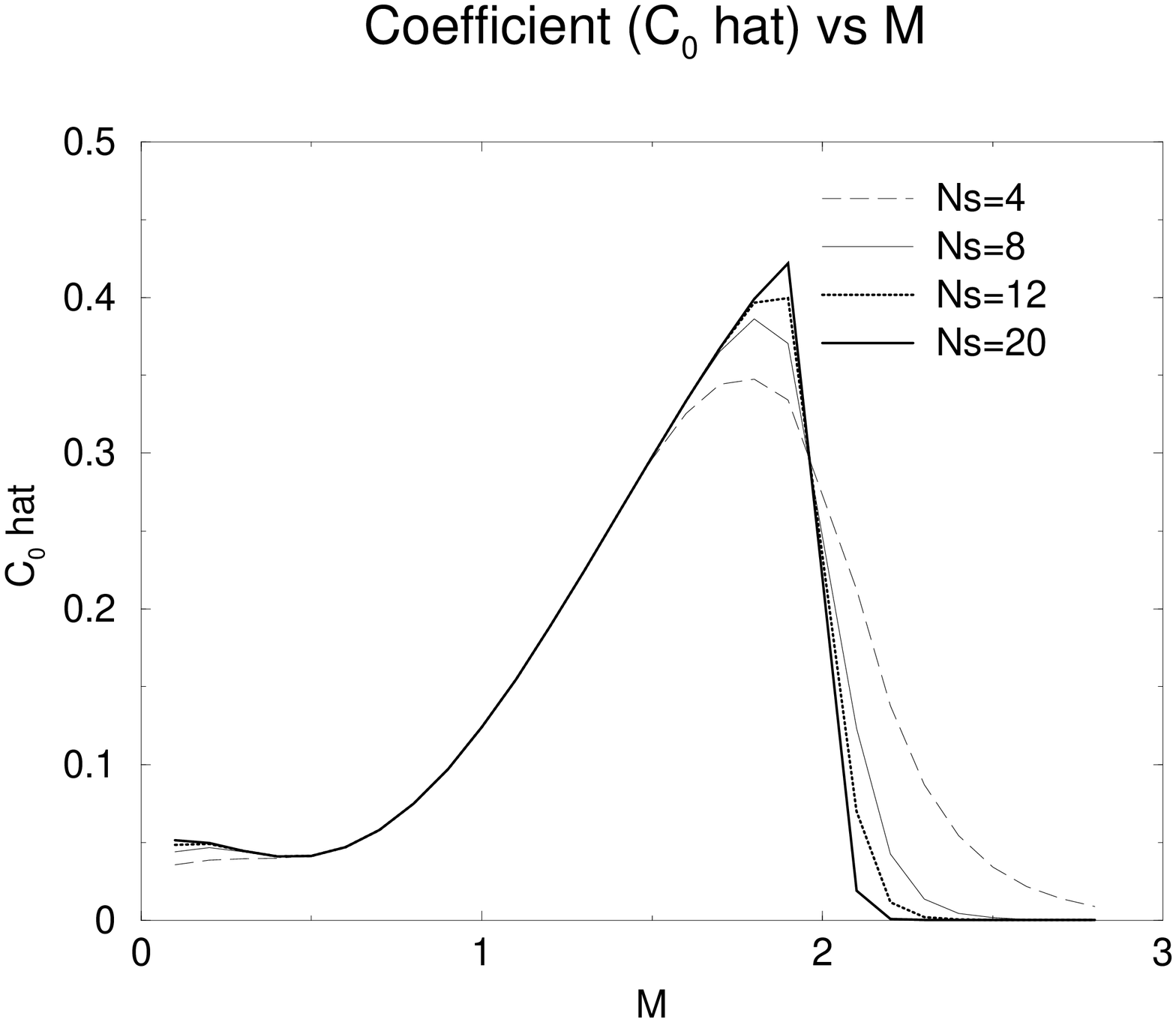}}
\caption{
The coefficient, $\hat{C_0}$, as a function of $M$
for  various $N_s$.
}
\label{fig:coef_c0}
\end{figure}

$I_0$ is calculated in the same way as in $N_s=\infty$ case:
\begin{equation}
I_0 = \int_{0}^{{\sigma^{\prime}}^2+\Pi^2} d \rho K(\rho)
\quad , \quad 
K(\rho) = \int_{-\pi/a}^{\pi/a} \frac{d^2 p}{(2 \pi)^2}
\frac{1}{\frac{\displaystyle{H^{\prime}}}{\displaystyle{F
a^2}} + \rho} \, .
\label{eq:kernel4}
\end{equation}
Here we take the divergent part in $a \rightarrow 0$.
The shifted function $H'$ has similar behavior to the limit $N_s\to\infty$:
\begin{equation}
\frac{H^\prime}{F a^2} \rightarrow
f_M^{-2} \sum_{\mu} p_\mu^2  ,
\end{equation}
with 
\begin{equation}
f_M = \frac{M (2-M)}{1 - (1-M)^{2 N_s}} \, .
\label{eq:factor1}
\end{equation}
Therefore the logarithmic divergent term has similar form 
as continuum theory:
\begin{equation}
I_0 = - \frac{f_M^2}{4 \pi} \left({\sigma^{\prime}}^2 + \Pi^2 \right)
\ln \frac{a^2 f_M^2 \left({\sigma^{\prime}}^2 + \Pi^2 \right)}{e}
+ \hat{C_0} \left({\sigma^{\prime}}^2 + \Pi^2 \right) , 
\end{equation}
where the new constant $\hat{C_0}$ is defined by 
$\hat{C_0} = C_0 + C_0^{\prime}$ with
\begin{equation}
C_0(M,N_s) = \int_{-\pi}^{\pi} \frac{d^2 \xi}{(2 \pi)^2}
\frac{f_M^{-2} \sum_\mu \xi_\mu^2 - 
\frac{\displaystyle{H^\prime}}{\displaystyle{F}}}
{\frac{\displaystyle{H^{\prime}}}{\displaystyle{F}}
f_M^{-2} \sum_\mu \xi_\mu^2 } \, .
\end{equation}
and
\begin{equation}
\int_{-\pi/a}^{\pi/a} \frac{d^2 p}{(2 \pi)^2}
\frac{1}{f_M^{-2} \sum_{\mu=1}^{2} p_\mu^2 + \rho} \equiv
\frac{f_M^2}{4 \pi} \ln \frac{1}{a^2 f_M^2 \rho} +
C_0^{\prime} .(M,N_s) ~.
\end{equation}
The coefficient $\hat{C}_0$ as a function of $M$ for various $N_s$
is plotted in Fig.\ref{fig:coef_c0}.

\begin{figure}
\centerline{\epsfxsize=7.5cm \epsfbox{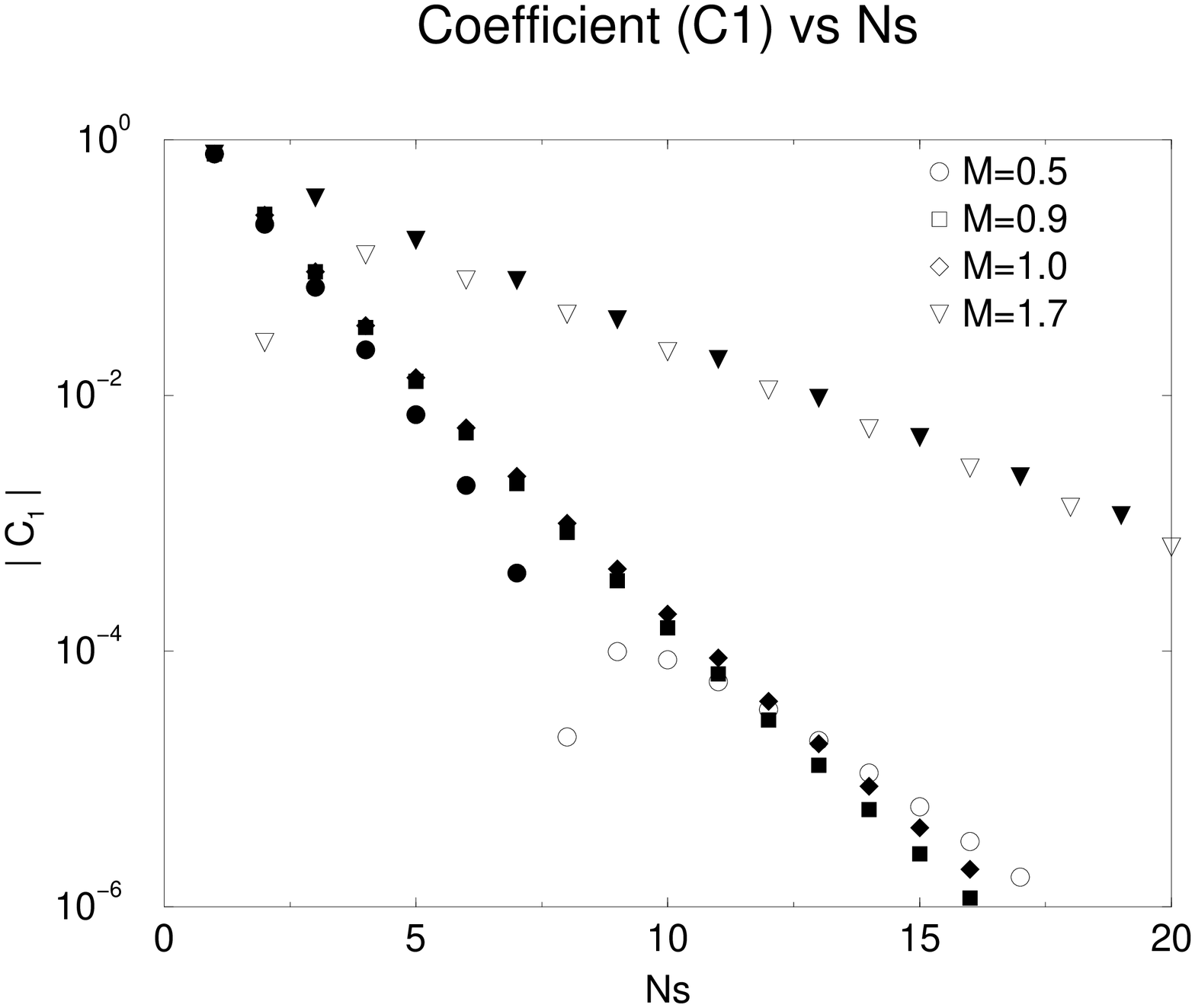}\ \
            \epsfxsize=7.5cm \epsfbox{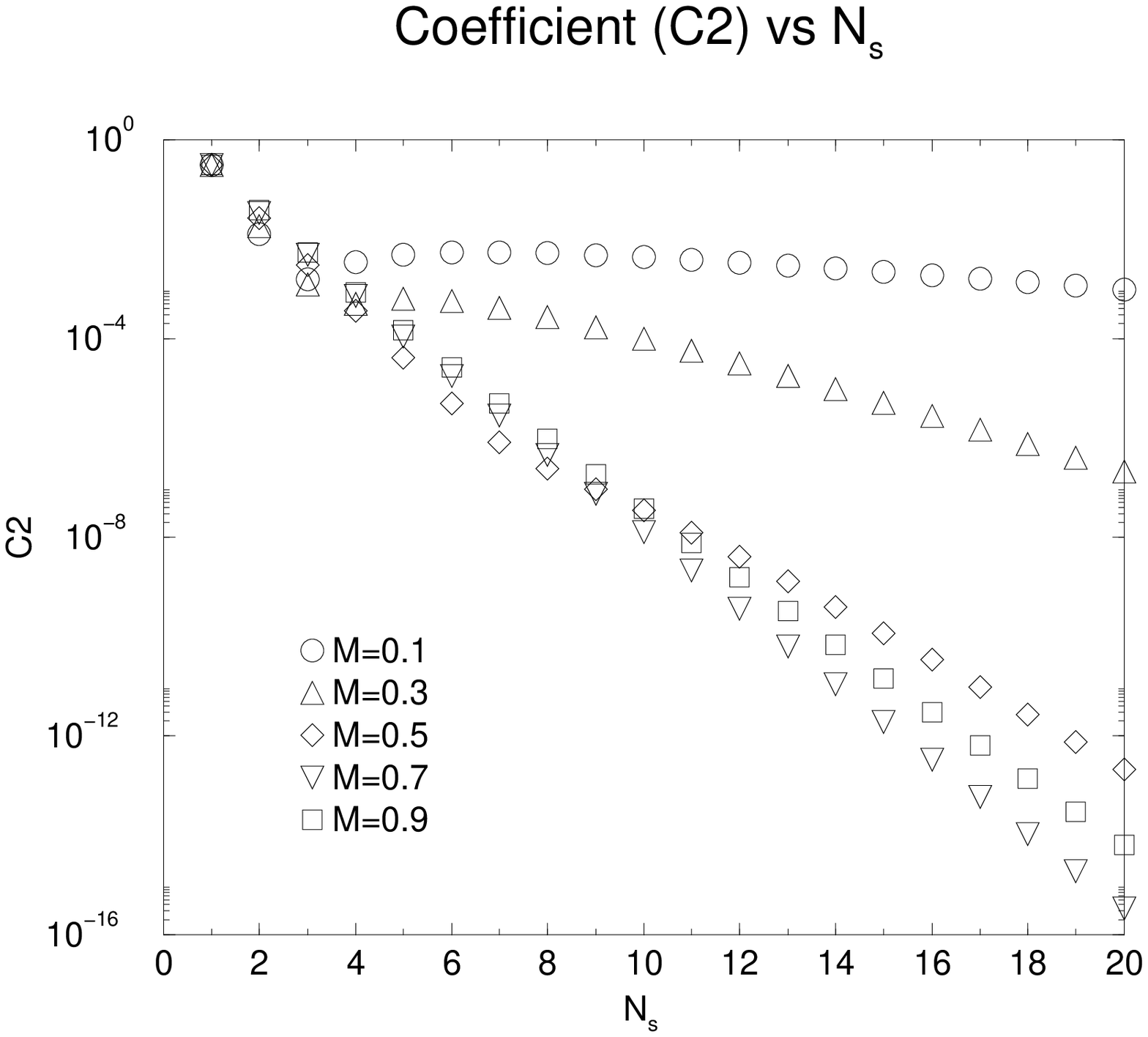}}
\caption{
The coefficients, $C_1$ and $C_2$, as a function of $N_s$
for fixed values of $M$. 
The data for $C_1$ is plotted absolute values and 
filled symbols denote  $C_1 < 0$.
}
\label{fig:coef_c1}
\end{figure}

Remaining $I_n$ in the continuum limit is easily calculated.
$I_1$ ($I_2$) is a linear divergent (constant) term 
in $a\to 0$ limit,
while $I_n$ ($n \geq 3$) vanishes:
\begin{eqnarray}
&&I_1 = \frac{\sigma^{\prime}}{a} C_1 
= \frac{\sigma^{\prime}}{a}  \int_{-\pi}^{\pi} 
\frac{d^2 \xi}{(2 \pi)^2} \frac{G^{\prime}}{H^{\prime}}
\, ,\\
&&I_2 = - {\sigma^{\prime}}^2 C_2
= - {\sigma^{\prime}}^2   \frac{1}{2} 
\int_{-\pi}^{\pi}  \frac{d^2 \xi}{(2 \pi)^2} 
\left[ \frac{G^{\prime}}{H^{\prime}} \right]^2 \, .
\label{eq:C2}
\end{eqnarray}
Fig.\ref{fig:coef_c1} shows these coefficients
as a function of $N_s$ for various $M$.
 
From above calculations,  
the effective potential in finite $N_s$ case in the continuum limit
is obtained as follows:
\begin{eqnarray}
V_{eff} &=& - \left( \frac{m_f^{\prime}}{g_\sigma^2} +
\frac{C_1}{a} \right) \frac{1}{f_M} \sigma_R
+ \left(\frac{1}{2 g_\sigma^2} - \hat{C_0} + C_2 \right)
\frac{1}{f_M^2} \sigma_R^2 \nonumber \\
&+& \left(\frac{1}{2 g_\pi^2} - \hat{C_0} \right) 
\frac{1}{f_M^2}\Pi_R^2 
+ \frac{1}{4 \pi} \left(\sigma_R^2 + \Pi_R^2 \right) 
\ln \frac{a^2 \left(\sigma_R^2 + \Pi_R^2 \right)}
{e} , 
\label{eq:effpot1}
\end{eqnarray}
where
\begin{equation}
m_f^\prime = m_f - \frac{(1-M)^{N_s}}{a} \quad , \quad 
\sigma_R = f_M \sigma^\prime \quad , \quad 
\Pi_R = f_M \Pi \, .
\label{eq:MfShift}
\end{equation}
If the coupling constants and the mass parameter are renormalized as
\begin{eqnarray}
&&\frac{1}{2 g_\sigma^2} - \hat{C_0} + C_2 
= \frac{f_M^2}{4 \pi} \ln \frac{1}{a^2 \Lambda^2} , 
\label{eq:gsgtuning}\\
&&\frac{1}{2 g_\pi^2} - \hat{C_0}  
= \frac{f_M^2}{4 \pi} \ln \frac{1}{a^2 \Lambda^2} , 
\label{eq:gpituning}\\
&&\frac{1}{f_M} \left( \frac{m_f^\prime}{g_\sigma^2} 
+ \frac{C_1}{a} \right) = m ~,
\label{eq:tuning}
\end{eqnarray}
we obtain the correct effective potential of the continuum theory 
(\ref{eq:contVeff}).

We emphasize that the current quark mass term, $m_f$, 
receives the $O(1/a)$ additive renormalization for $N_s=$ finite case.
As we describe previously,
fine tuning for bare mass parameter in (\ref{eq:tuning}) is 
needed for finite $N_s$, and chiral symmetry is restored 
in the continuum limit.

The restoration of chiral symmetry in $N_s\to\infty$ could also be seen
in the scaling relation (\ref{eq:gsgtuning}-\ref{eq:tuning}).
The coefficients $C_1$ and $C_2$ represent the magnitudes of 
the explicit breaking of chiral symmetry. 
$C_1$ is the additive mass counter term, 
while $C_2$ is the miss match between the quadratic terms 
of scalar and pseudo scalar particle
by the quantum correction.
If we restrict DW-mass to $0<M<2$, 
as shown in Fig.\ref{fig:coef_c1},
the coefficients $C_1$ and $C_2$ are decreased rapidly
as $N_s$ becomes large.
The effects of shifting of the fields 
and the additive renormalization of 
the mass parameter in (\ref{eq:MfShift}) and (\ref{eq:tuning})
rapidly vanishes with increasing $N_s$,
thus the necessity of fine tuning becomes absent in $N_s\to\infty$ limit.

On the other hand, even when $M$ is set out of the region $(0,2)$,
the scaling relations (\ref{eq:gsgtuning}-\ref{eq:tuning}) lead
the chiral symmetric continuum limit. 
In this case $C_1$ and $C_2$ don't vanish in large $N_s$ limit 
and the fine tuning is necessary, similar to WF.

\section{The parity broken phase and the restoration of chiral symmetry}
\label{sec:Aokiphase}

For lattice QCD with WF, 
the existence of massless pion 
for finite lattice spacing is explained by the parity broken phase 
picture proposed by Aoki \cite{aoki}.
Although chiral symmetry is explicitly broken in WF,
one can tune the mass parameter to obtain
an exact {\it massless} pion in the spectrum even 
for finite lattice spacing.
This cannot be understood by the ordinary picture of Nambu-Goldstone boson
in the continuum theory, in which  chiral symmetry is the exact 
symmetry of the action and is broken spontaneously.

Aoki examined the GN model and lattice QCD with WF
for a finite lattice spacing, 
and found that the parity symmetric phase and 
parity(--flavor) spontaneously broken phase 
coexist in the parameter space of the model.
The parity broken phase is characterized by the non-zero condensation of 
the pseudo scalar density,
$\langle\Pi\rangle=\langle\psi i\gamma_5\psi\rangle\neq0$.
Provided a second-order phase transition separating the two phases 
from each other, pion becomes massless at the phase transition point,
which is regarded as  a massless particle  accompanying with 
the continuous phase transition.

Before starting analysis of phase diagram of DWF model for  general $N_s$,
we note the equivalence between $N_s=1$ DWF model 
and Wilson fermion formalism.
The effective potential of WF,
\begin{eqnarray}
&& V_{\rm W} = \frac{1}{2 g_{\sigma}^2} \left( \sigma - m_f \right)^2
 +  \frac{1}{2 g_{\pi}^2} \Pi^2 \nonumber \\
&&-\int_{-\pi/a}^{\pi/a} \frac{d^2 k}{(2 \pi)^2}
\ln \left[ \sum_\mu \frac{\sin^2 k_\mu a}{a^2} + \left
( \sigma_c + \frac{r}{a} \sum_\mu (1 - \cos k_\mu a)
\right)^2 + \Pi_c^2 
\right] \, ,
\label{eq:wilson}
\end{eqnarray} 
can be seen easily by substituting $N_s=1, \sigma_c=\sigma-(1-M)$ 
in (\ref{eq:pot}).
The phase boundary of  Aoki phase forms the three cusps which
reach the weak-coupling limit $g^2=0$ at $M=1,3,5$
in WF\cite{aoki,umemura,kaneda} and $N_s=1$ DWF model.  
Three cusps arise from the fact that the doublers
at the conventional continuum limit, $(g^2,M)=(0,1)$, become physical 
massless modes at at $M=3,5$.

On the other hand, in $N_s=\infty$ limit, the parity broken phase 
does not exist because of the restoration of chiral symmetry, 
whose explicit breaking causes the parity broken phase.

For $1<N_s<\infty$, it is expected that
the parity broken phase exists and the chiral phase boundary forms the cusps
as surmised from the Wilson like behavior in the previous section.
The region of the broken phase is distorted and shrinks rapidly 
as $N_s$ is increased, and vanishes in $N_s \to \infty$ limit.

The parity symmetry is spontaneously broken at 
parameter point $(m_f,g^2; M,N_s)$, 
where the gap equations have a stable solution $\Pi\neq0$.
Setting the  parameters of this model 
$(m_f,g_\sigma^2,g_\pi^2, M,N_s)$ ,
observables are calculated by solving gap equations,
\begin{eqnarray}
&&\frac{\partial V_{eff}}{\partial \sigma} = 
\frac{1}{g_\sigma^2} \left(\sigma - m_f \right) 
- \sigma {\cal {F}}(\sigma , \Pi) 
- {1\over a} {\cal {G}}(\sigma,\Pi)
= 0 \, , 
\label{eq:gapsg} \\
&&\frac{\partial V_{eff}}{\partial \Pi} = 
\Pi \left[ \frac{1}{g_\pi^2} 
- {\cal {F}}(\sigma , \Pi) \right]
= 0  \, ,
\label{eq:gappi} 
\end{eqnarray}
where
\begin{eqnarray}
&&{\cal{F}}(\sigma,\Pi) =  \int_{-\pi}^{\pi} \frac{d^2 \xi}{(2 \pi)^2}
\frac{ 2 F }{H + G a \sigma + Fa^2(\sigma^2 + \Pi^2) } \, , \\
&&{\cal{G}}(\sigma,\Pi) = \int_{-\pi}^{\pi} \frac{d^2 \xi}{(2 \pi)^2}
\frac{ G }{H + G a \sigma + F a^2(\sigma^2 + \Pi^2) } \, , 
\end{eqnarray}
with $\xi_\mu = p_\mu a$.
In the case of $g_\pi^2=g_\sigma^2=g^2$,
the position of the phase boundary in the parameter space
could be obtained by solving these gap equations for $\Pi=\epsilon$.
Taking $\epsilon\to0$ limit, 
(\ref{eq:gapsg}) (\ref{eq:gappi}) define $g^2$ and $m_f$ as
functions of $\sigma$ which is nothing but 
the parametric representation of the phase boundary  
if the phase transition is continuous.

Let us note that DWGN model with $g^2=g_\sigma^2=g_\pi^2$ has only
second order phase transition between the parity symmetric and 
the broken phases. 
This is in contrast with WF case \cite{FiniteMuGN},
in which theory a first order phase transition 
for $g_\sigma^2 \neq g_\pi^2$ is found.

By differentiating (\ref{eq:gappi}) with respect to $\Pi$,
one can easily check from that
pion mass squared 
\begin{equation}
M_\pi^2 \propto \widetilde{M}_\pi^2 \equiv {1\over f_M^2}
{\partial^2 V_{eff}(\sigma,\Pi=0) \over \partial\Pi^2} 
= {1\over f_M^2} \left[ {1 \over g^2 }- {\cal F}(\sigma,\Pi=0) \right] ~,
\label{eq:PiMass}
\end{equation}
exactly vanishes at the phase boundary.
The factor, $1/f_M^2$, in front of the r.h.s corrects the
normalization of auxiliary field $\Pi$ in the same sense as 
explained in the previous section.
(See below (\ref{eq:coef1}).)

\begin{figure}
\centerline{\epsfxsize=10cm \epsfbox{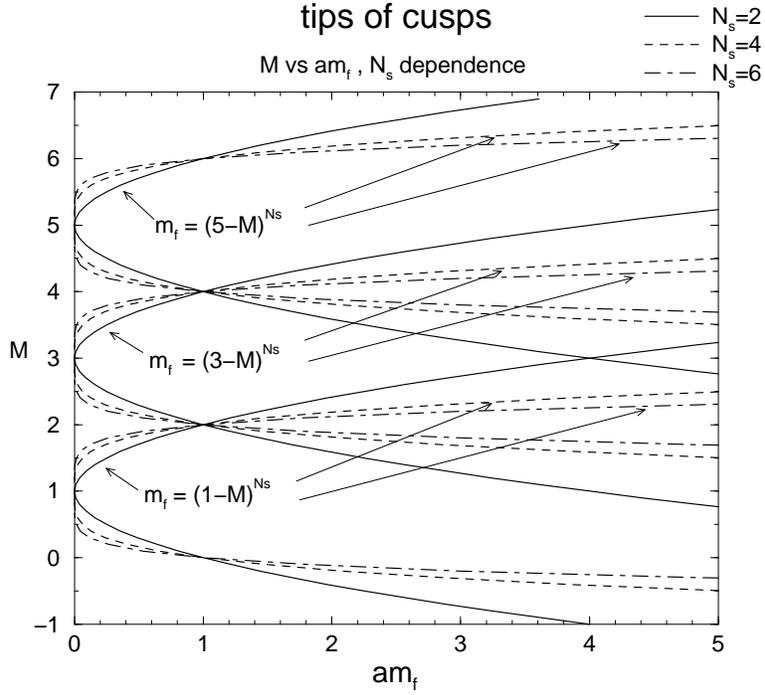}}
\caption{
The chiral symmetric continuum limit in $(M,am_f)$ plane
for $N_s$=even.
The critical lines approach to $m_f=0$  with increasing $N_s$ in 
each of three region, $2l<M<2(l+1), l=0,1,2$.
}
\label{fig:tip_e}
\end{figure}

Before drawing whole phase diagrams in the parameter space 
$(m_f, g^2)$, let us first examine the positions of 
the three ``tips of cusps", which are the intersection points 
between $g^2=0$ plane and the critical line.
These tips correspond to the continuum limits $g^2\to 0$ 
similar to WF.
The positions can be obtained by the asymptotic form
of the  integrals in (\ref{eq:gapsg}) and (\ref{eq:gappi}). 
The divergent behaviors of the integrals around 
$ p_\mu a =(\pi m,\pi n)$,
with $m , n=0 \,\,\, \mbox{or} \,\,\, 1$,
are
\begin{eqnarray}
{\cal{F}}(\sigma,\Pi=0) &\rightarrow&
\sum_{m,n=0,1,l=m+n} \int_{-\pi}^\pi \frac{d^2 \xi}{(2 \pi)^2}
\frac{2}{\left\{ \sigma - (1-M + 2 l)^{N_s}
\right\}^2 + \sum_\mu {\cal{A_\mu}} \xi_\mu^2} \, ,
\label{eq:funcFasy} \\
{\cal{G}}(\sigma,\Pi=0) &\rightarrow&
\sum_{m,n=0,1,l=m+n} \int_{-\pi}^\pi \frac{d^2 \xi}{(2 \pi)^2}
\frac{-2 (1-M+2 l)^{N_s}}{
\left\{ \sigma  - (1-M + 2 l)^{N_s} \right\}^2 
+ \sum_\mu {\cal{B_\mu}} \xi_\mu^2} \, ,
\label{eq:funcGasy}
\end{eqnarray}
where ${\cal{A_\mu}}$ and ${\cal{B_\mu}}$ are functions 
of $l, \sigma , M$ and $N_s$.
From these expressions it is easy to see that the r.h.s of the gap equations
(\ref{eq:gapsg}) (\ref{eq:gappi}) have a logarithmic divergence 
at $\sigma=(1 - M + 2 l)^{N_s}$ with $l=0,1,2$.
This fact leads that for each $l$ the phase boundary 
intercepts with $g^2 = 0$ at a point $m_f = (1 - M + 2 l)^{N_s}$.
Each of three critical points corresponds to the massless particle pole 
of momentum 
\begin{equation}
p_\mu(l=0)=(0,0) \quad , \quad p_\mu(l=1)= (\pi/a,0),(0, \pi/a)
\quad , \quad p_\mu(l=2)= (\pi/a,\pi/a) ~.
\label{eq:ThreeMoms}
\end{equation}
The positions of three points move as a function of $M$, 
which is shown in  Fig.\ref{fig:tip_e}.
For $M$ in the region of $(2l, 2(l+1))$, 
the chiral point for momentum mode $p_\mu(l)$ in (\ref{eq:ThreeMoms}) 
converges to $m_f=0$  as $N_s$ increases. 
Thus massless pion  can be automatically obtained at $m_f=0$ 
in the limit of $g^2=0$ by increasing $N_s$.
Other two critical points move rapidly to $|m_f|\to\infty$ in 
$N_s \to \infty$ limit.

\begin{figure}
\centerline{\epsfxsize=10cm \epsfbox{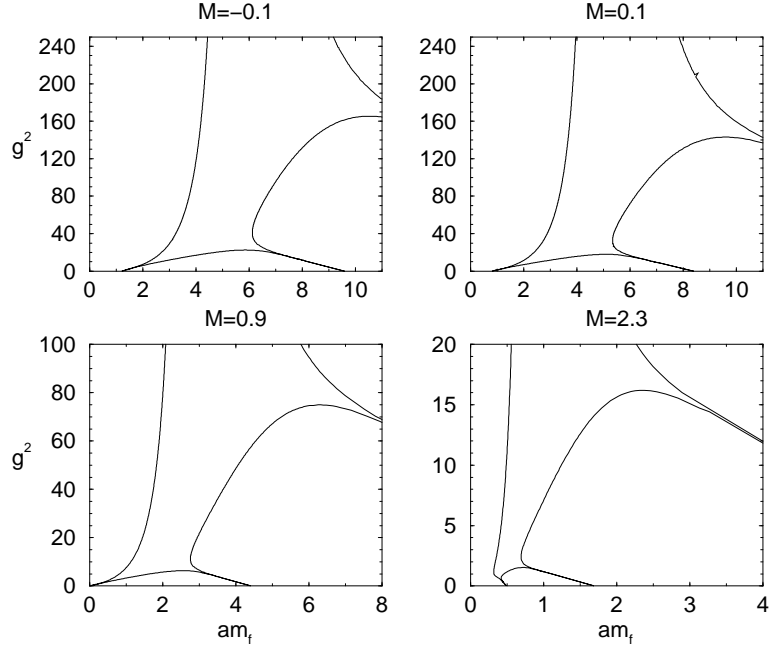}}
\caption{
The phase diagram at $N_s = 2$. 
The horizontal axis represents $a m_f$ while the vertical axis is $g^2$.
Each of four graphs is plotted for fixed value of $M$.
The inside region of the critical line is the parity broken phase.
}
\label{fig:Ns2}
\end{figure}

\begin{figure}
\centerline{\epsfxsize=10cm \epsfbox{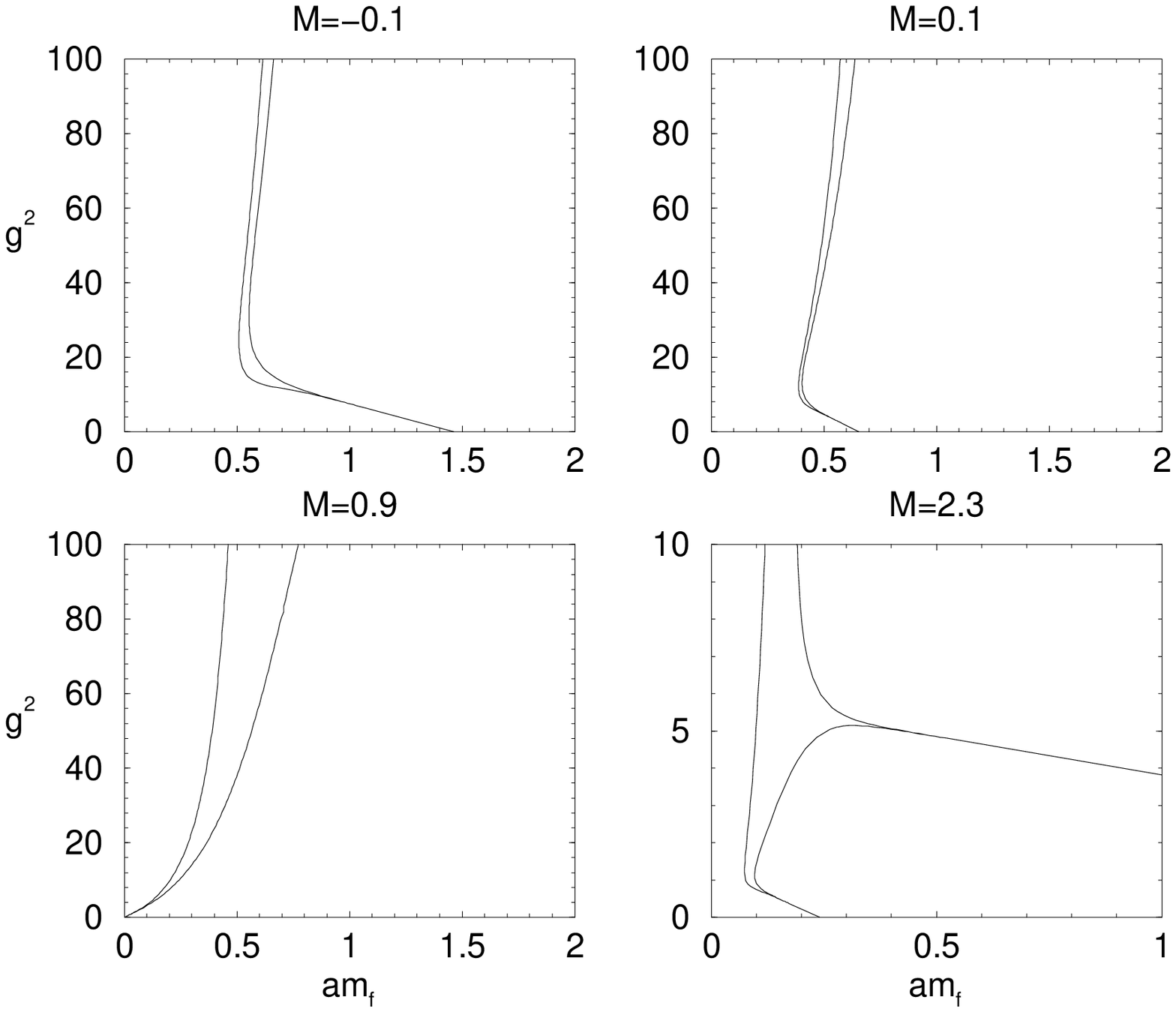}}
\caption{
The phase diagrams at $N_s=4$.
}
\label{fig:Ns4}
\end{figure}

\begin{figure}
\centerline{\epsfxsize=10cm \epsfbox{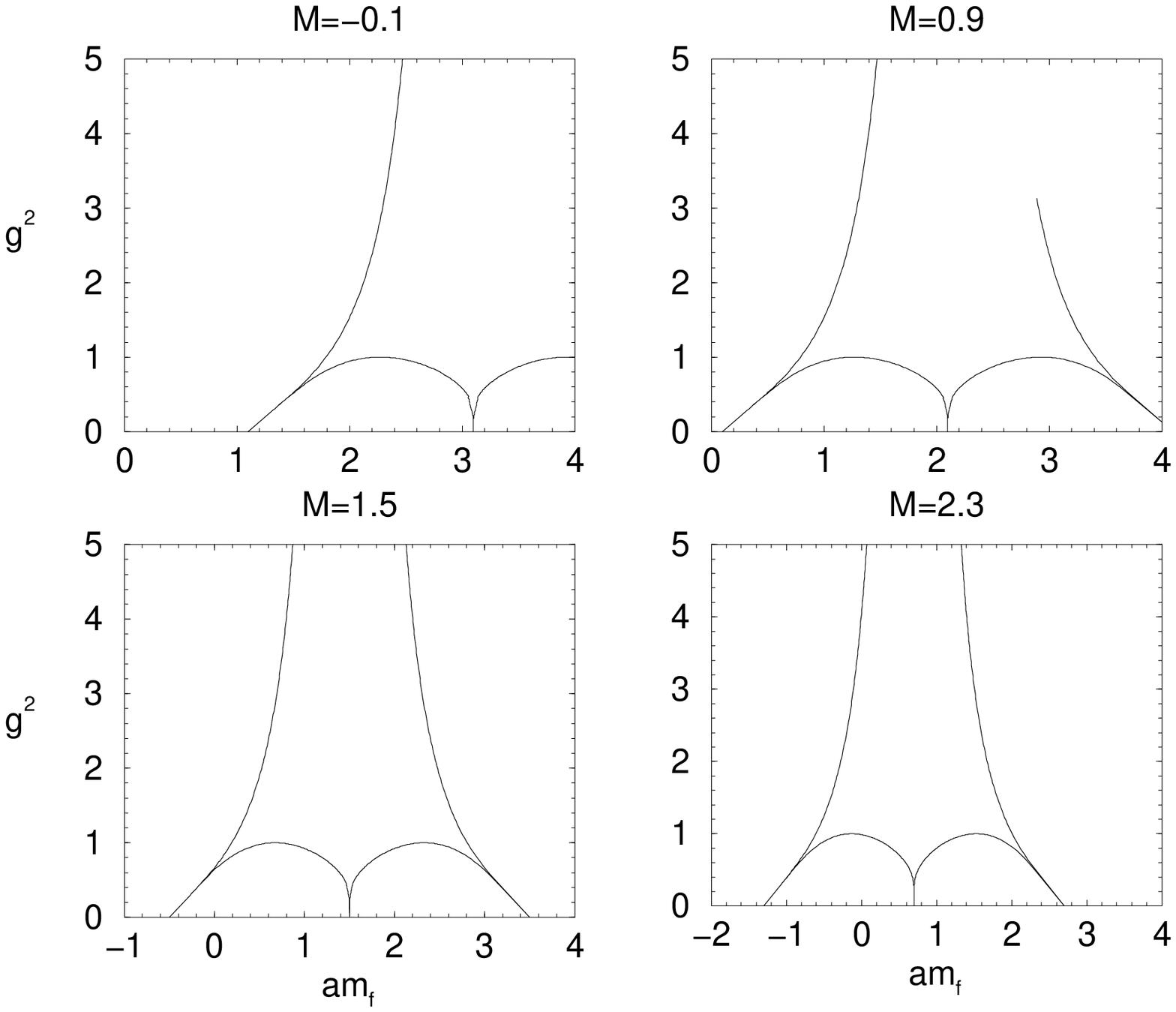}}
\caption{
The phase diagrams at $N_s=1$.
}
\label{fig:Ns1}
\end{figure}

\begin{figure}
\centerline{\epsfxsize=10cm \epsfbox{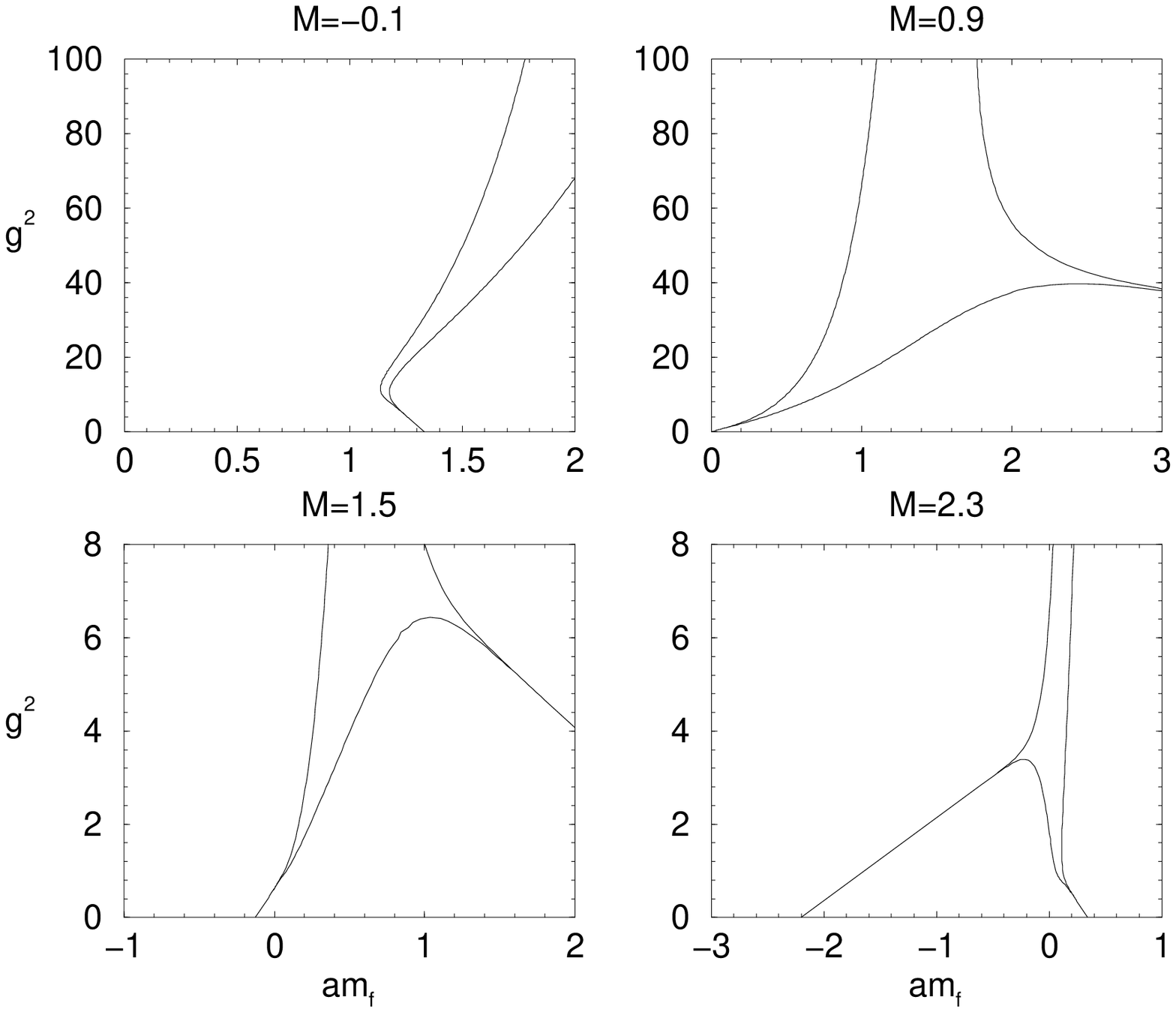}}
\caption{
The phase diagrams at $N_s=3$.
}
\label{fig:Ns3}
\end{figure}

The actual calculation for the chiral phase boundary 
is done by solving the gap equations numerically,
and results are shown in Figs.\ref{fig:Ns2}, \ref{fig:Ns4},
\ref{fig:Ns1} and  \ref{fig:Ns3}.
The former two figures are for $N_s=$ even cases
and the latter two show $N_s=$ odd cases.

\begin{figure}
\centerline{\epsfxsize=10cm \epsfbox{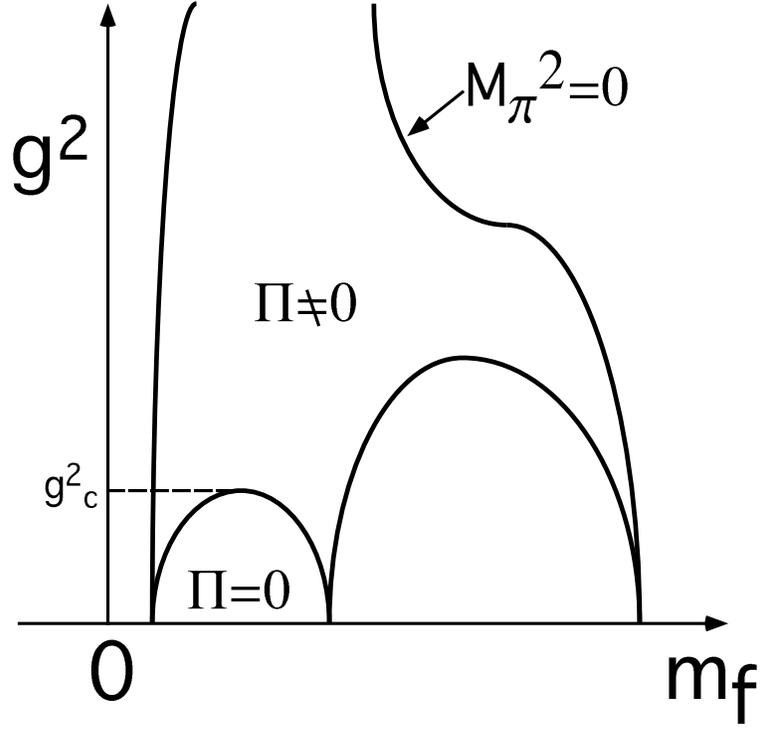}}
\caption{
The schematic phase diagram 
in $(m_f,g^2)$ plane for $N_s=$even.
The parity broken phase exists in $m_f > 0$ region.
}
\label{fig:illust1}
\end{figure}

\begin{figure}
\centerline{\epsfxsize=9cm \epsfbox{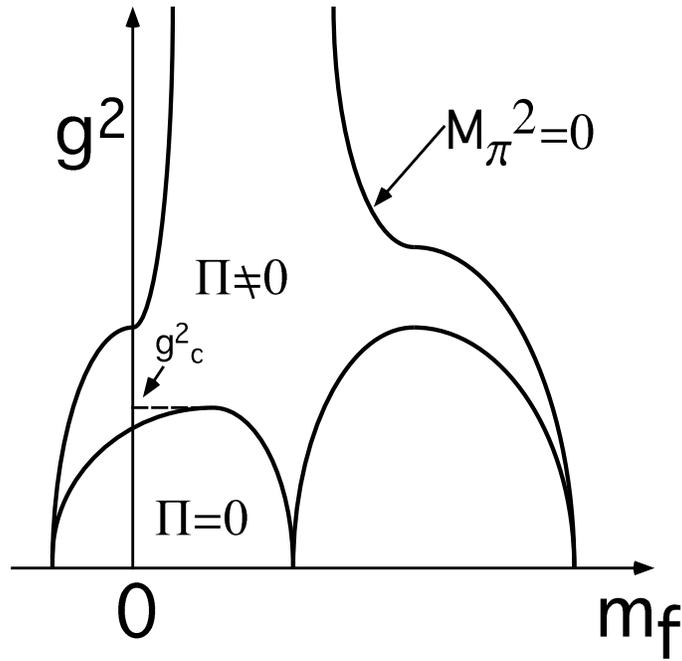}}
\caption{
Same as 
Fig.\ref{fig:illust1} 
for $N_s=$odd.
The parity broken phase lies across $m_f=0$ line.
}
\label{fig:illust2}
\end{figure}

The schematic diagram of the phase boundary 
between the parity symmetric and broken phase 
in $(m_f,g^2)$ plane for fixed $(M, N_s)$ 
is drawn in Fig.\ref{fig:illust1}
for $N_s=$even case
and in Fig.\ref{fig:illust2} for $N_s=$odd.
The phase boundaries in both cases  show queer shapes 
which have three intercept points, 
$m_f=(1-M+2 l)^{N_s}$ with $l=0,1,2$, on $g^2=0$ line. 

We find that the phase diagram for $N_s=$even 
is different from that for $N_s=$odd.
In $N_s=$even case, one can analytically verify from the 
sign-definiteness of integrands of the gap equations that 
the phase boundary exists only for $m_f > 0$ region.
(For the notation in Ref.\cite{Vra} and 
numerical simulations, the broken phase appear in $m_f<0$ region.)
On the other hand, the phase boundary intersects with $m_f=0$ line
in $N_s=$odd case. 
If $g^2$ is  decreased from finite value to zero,
the phase boundary moves from positive $m_f$ region to
$m_f = (1 - M + 2 l)^{N_s}$ at $g^2=0$,
which becomes negative in some region of $M$ for $N_s=$odd.
The chiral phase boundaries for $N_s=$ odd are shown
in Figs. \ref{fig:Ns1} and \ref{fig:Ns3}.
The parity broken phase lies across $m_f=0$ plane 
in $N_s=$odd case.
As similar for $N_s=$even,
this boundary for odd $N_s$ 
also converges to $m_f=0$ plane for larger $N_s$.

\begin{figure}
\centerline{\epsfxsize=12cm \epsfbox{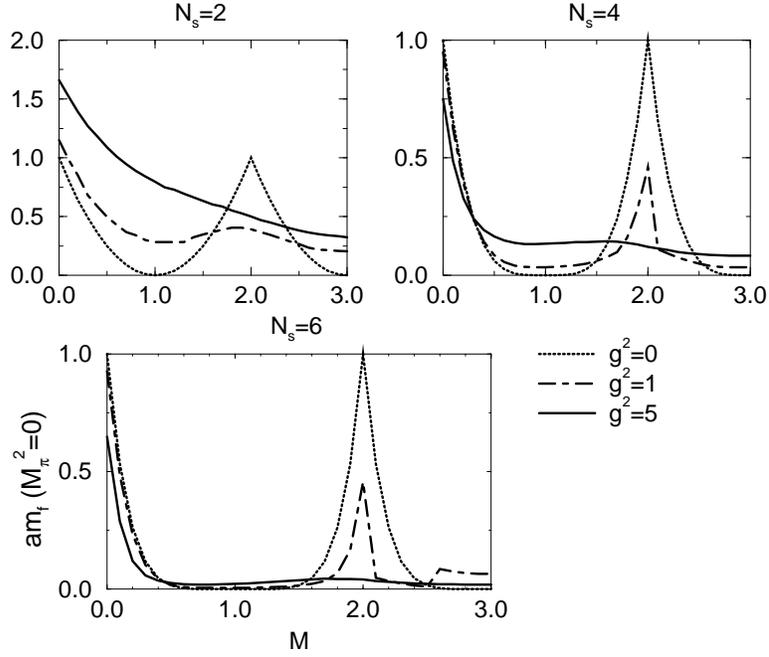}}
\caption{
The distance from $m_f=0$ line to the phase boundary;$m_f(M_\pi=0)$, 
which corresponds to the critical value of
current quark mass,
as a function of $M$ for fixed $N_s, g^2$.
}
\label{fig:mfM}
\end{figure}

\begin{figure}
\centerline{\epsfxsize=11.5cm \epsfbox{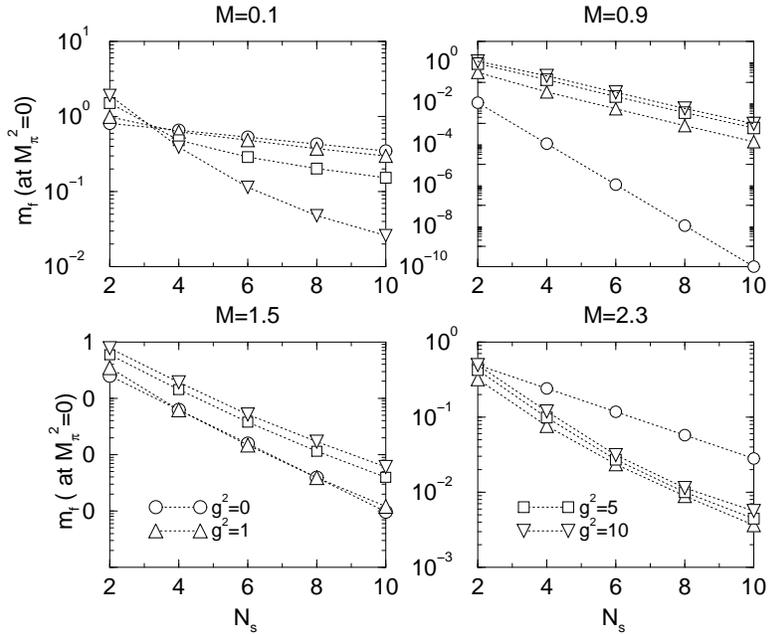}}
\caption{
$m_f(M_\pi=0)$ as a function of $N_s$ for fixed $M, g^2$.
}
\label{fig:mfNs}
\end{figure}

Another interesting observation from the phase diagram 
is finite $g^2$ effect for the pion mass.
The effective quark mass from the free propagator analysis 
(mean field analysis) is
\begin{equation}
m_q a \sim (M_c-M)^{N_s}~,
\label{eq:MFmass}
\end{equation}
from which one may think that there exists optimal $M (=M_c)$, where
the quark (and pion) becomes  massless even for finite $N_s$.
In current model, for finite $N_s$ the parity broken phase boundary
stays apart from $m_f=0$ plane {\it for all} $M$ if $g^2>0$.
This fact indicates that there is no massless pion at $m_f=0$ for 
finite $N_s$ and finite $g^2$.
We plot the value of $m_f$ where $M_\pi^2=0$ as a function of $M$ in
Fig. \ref{fig:mfM}.
(This is the distance of the phase boundary point from the $m_f=0$ line.
This quantity corresponds to the critical value 
of the inverse hopping parameter, $1/\kappa_c$, in WF.)
We find that for fixed finite $N_s$ and finite $g^2$, 
pion does not become massless at $m_f=0$ .  
We realize that this discrepancy from the mean field picture
in (\ref{eq:MFmass})
is due to the part of the additive mass term
which is proportional to  $C_1$ in (\ref{eq:tuning}).
For strong coupling, for example $g^2=5.0$,
$m_f(M_\pi=0)$ is almost flat at around $M \sim 2.0$.
This indicates a difficulty of detecting ``allowed region'' of $M$
by observing pion mass for strong coupling region.
For $g^2=0$ the theory turns out to be free theory, 
and $M=1$ gives $M_\pi=0$ at $m_f=0$.
If $N_s$ is increased,
$m_f(M_\pi=0)$ becomes 
exponentially small convergence to zero no matter 
whether $g^2$ is finite or 0 (See Fig.\ref{fig:mfNs}).

\begin{figure}
\centerline{\epsfxsize=8cm \epsfbox{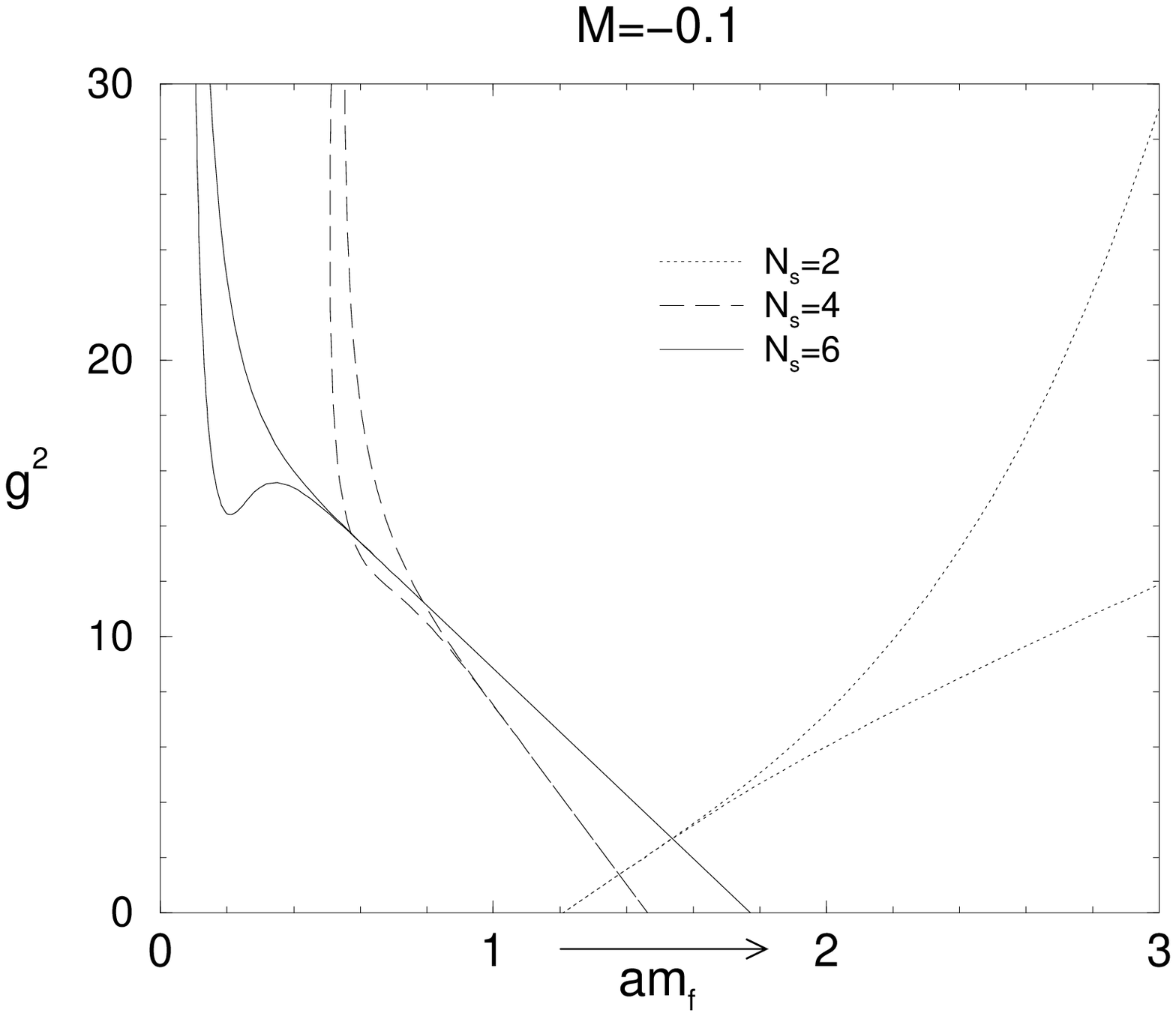}\ \ 
            \epsfxsize=8cm \epsfbox{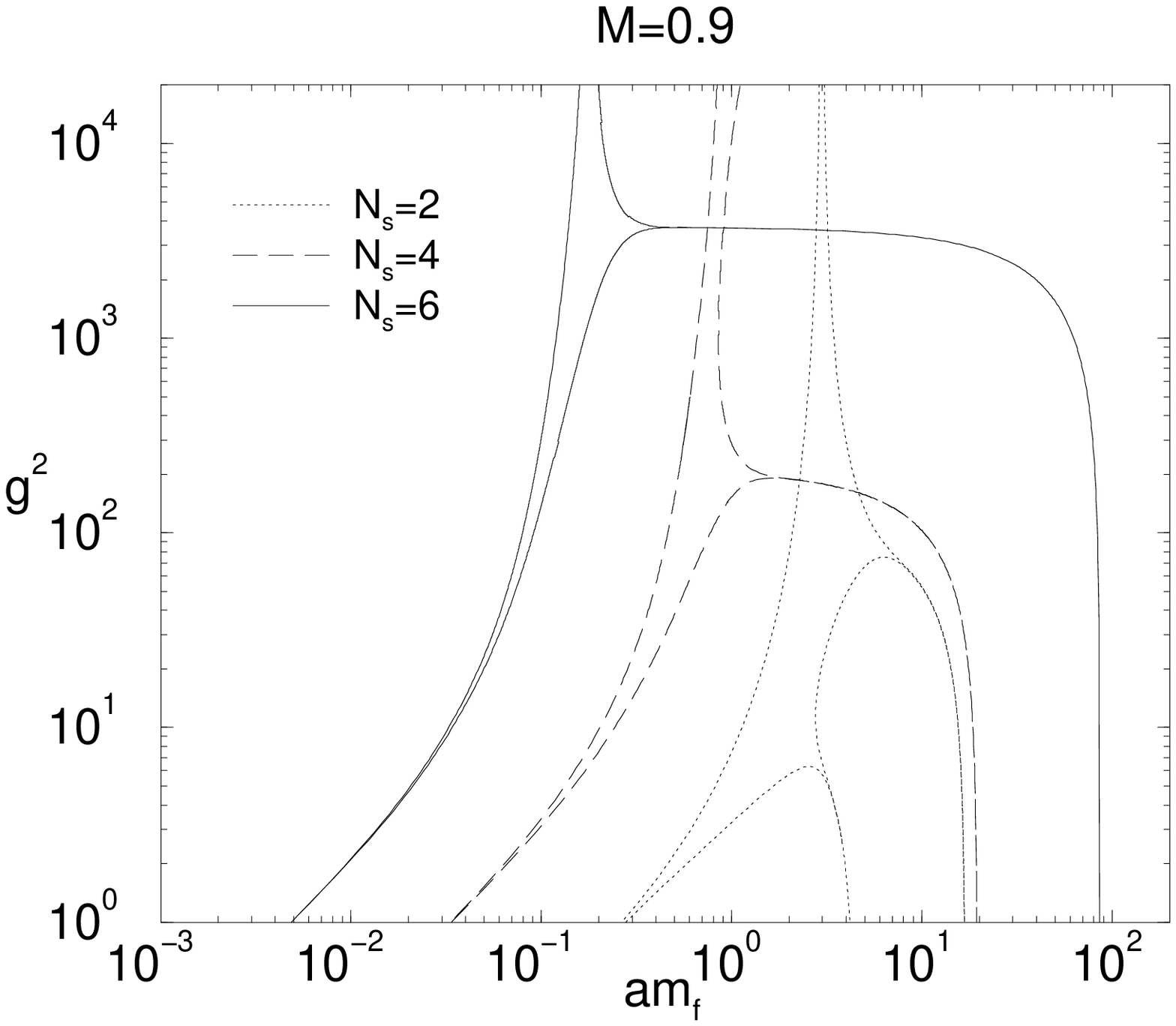}}
\caption{
$N_s$ dependence of the phase boundary for $M=-0.1 , 0.9$. 
The horizontal axis represents $m_f$ while 
the vertical axis is $g^2$.
}
\label{fig:MNs}
\end{figure}

The dependence of the phase diagram to $N_s$ could be seen in
Fig.\ref{fig:MNs}. 
In the figure for $M=0.9$,
the cusp of smaller $m_f$ corresponds to the chiral continuum limit of
the conventional momentum mode $p_\mu=(0,0)$
and another cusps show that of the doubler modes.
The first cusp converges to $m_f=0$ line while the other cusps
diverge to $m_f\to\infty$ in large $N_s$ limit.
Below a critical coupling, $g^2_c$, 
in Figs.\ref{fig:illust1} and \ref{fig:illust2},
a pair of critical lines forms
either of three continuum limits (three cusps) similar to WF, which
can be understood as the signal for the recovery of chiral symmetry.
For $g^2 > g^2_c$ the parity broken phases are merged 
to a uniform structure.
We find that 
$g^2_c$ increases exponentially with $N_s$;
\begin{equation}
g^2_c \sim e^{c N_s} ~, \quad (c > 0) ~.
\end{equation}
At the same time,
the phase boundary exponentially approaches to 
$m_f=0$ plane with increasing $N_s$.
For any $g^2$, the phase boundary converge at $m_f=0$ plane, 
which can be easily seen from (\ref{eq:gapsg}) and (\ref{eq:gappi}).
The width of the parity broken phase scales proportional to 
$a^3$\cite{phase}, and shrinks exponentially for $N_s\to\infty$ . 
these facts are compatible with the exact chiral symmetry for finite 
lattice spacing in $N_s\to\infty$ limit, as discussed in previous section. 

On the other hand,
at $M=-0.1$
the chiral continuum limit and the phase boundary 
go away from $m_f=0$ line with increasing $N_s$.
This behavior shows the violation of chiral symmetry
at $m_f=0$ even in $N_s \to \infty$ limit 
for $M<0$.

\section{$(M,N_s)$ dependencies of lattice observables}
\label{sec:paradep}

Let us  turns to discuss about the physical observables in
this model.
We choose $\sigma, \langle\bar q q\rangle$ and $\widetilde{M}_\pi$ 
as physical observables.

\begin{figure}
\centerline{\epsfxsize=10cm \epsfbox{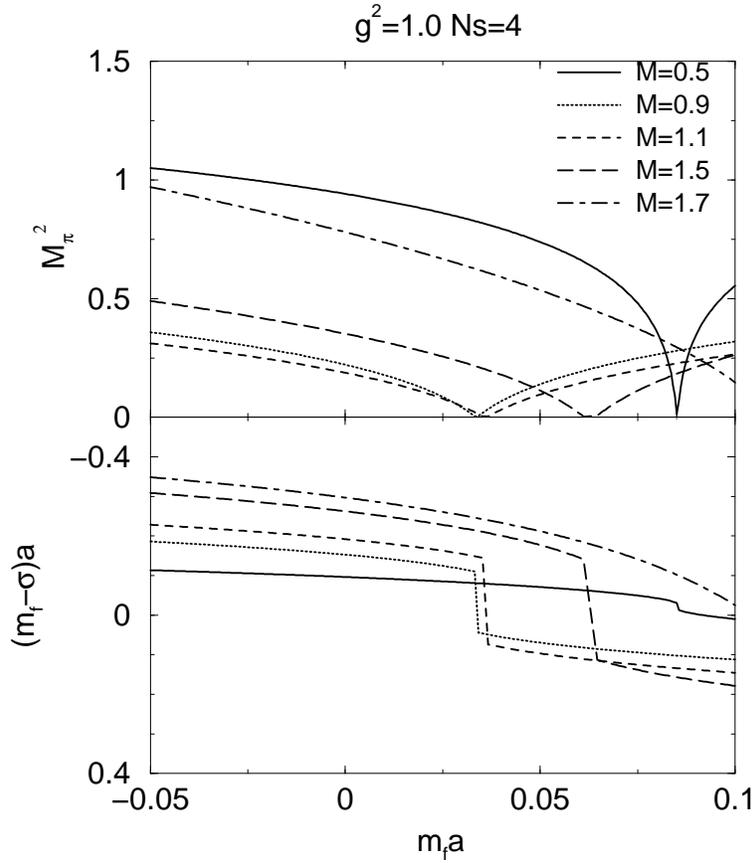}}
\caption{
$\widetilde{M}_\pi^2$ and 
$\langle \bar{q}q \rangle = m_f - \sigma$ as a function of $m_f$.
$N_s=4, g^2=1.0$.
}
\label{fig:Obs_mf-g1.0_Ns4}
\end{figure}

In Fig.\ref{fig:Obs_mf-g1.0_Ns4} 
we plot $\widetilde{M}_\pi^2$ and 
$\langle\bar q q\rangle\equiv m_f-\sigma$ as a function of $m_f$ for 
several values of $M$ with $g^2=1, N_s=4$. 
We find that there exists finite $m_f$ region in 
which $M_\pi$ is zero\footnote{
We set $\Pi=0$ throughout our calculation,
and the value of physical observables in the parity broken phase 
is not exact. 
For example, $M_\pi^2$ should not be zero in the parity
broken phase.
Note that all figures 
after this section except Fig.\ref{fig:Obs_mf-g1.0_Ns4}
are obtained in parity symmetric parameter region, 
in which our results are precise.}.
This region is nothing but the parity broken phase.
The $m_f$ dependence of the $\langle\bar q q\rangle$ shows a
discontinuous leap in between the parity broken phase.

\begin{figure}
\centerline{\epsfxsize=10cm \epsfbox{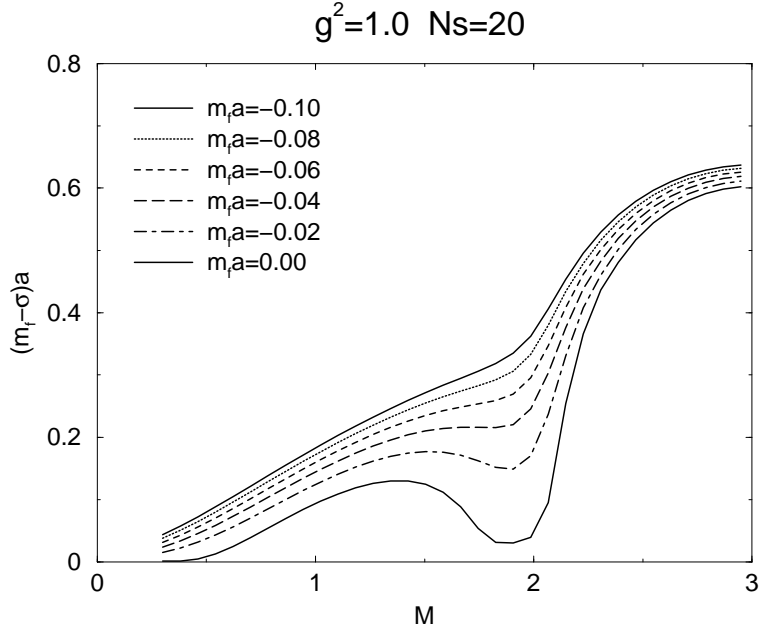}}
\caption{
$\langle \bar{q}q \rangle = m_f-\sigma$ as a function of $M$.
$N_s=20, g^2=1.0$.
}
\label{fig:pbp_M-g1.0}
\end{figure}

\begin{figure}
\centerline{\epsfxsize=7.5cm \epsfbox{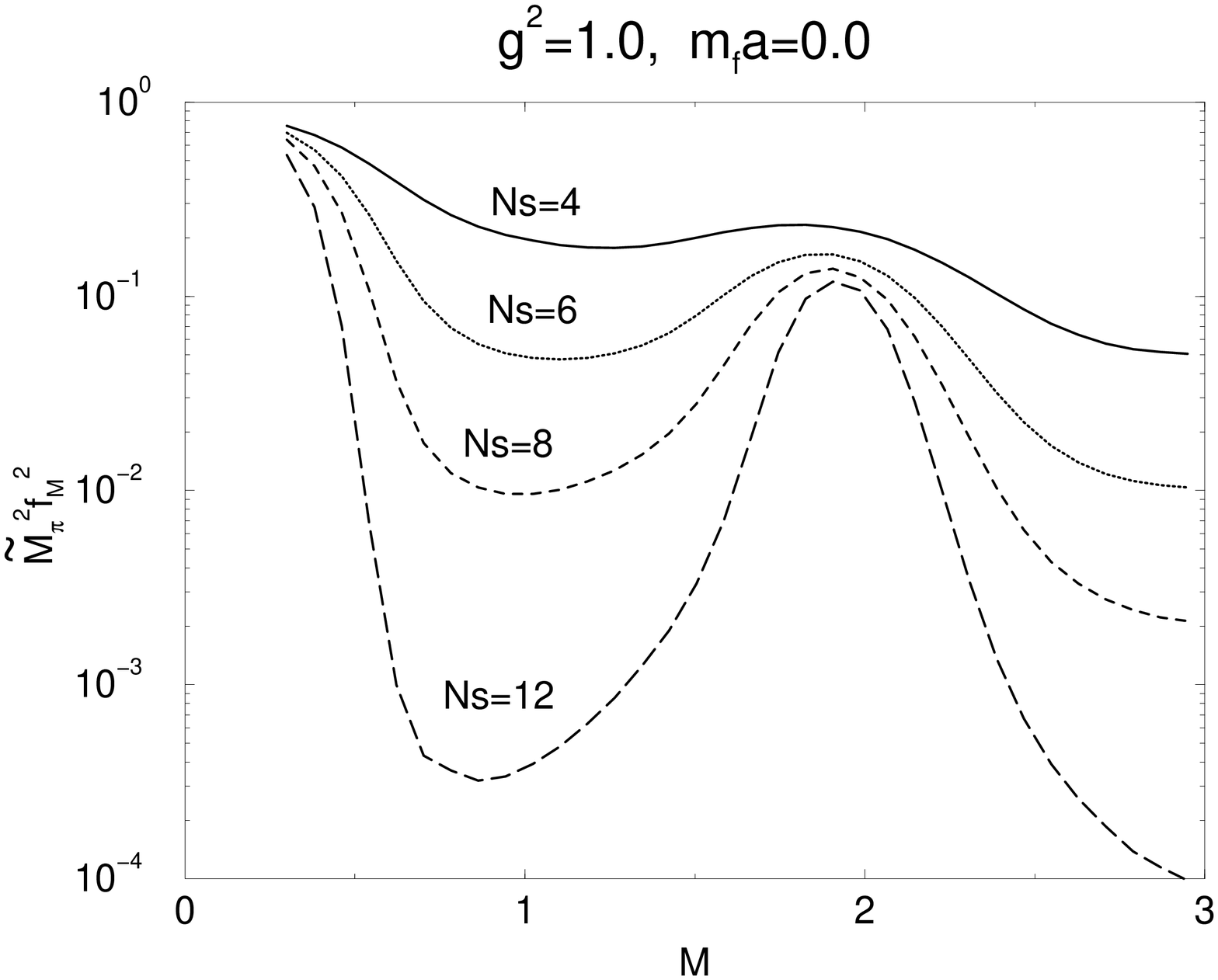}\ \
            \epsfxsize=7.5cm \epsfbox{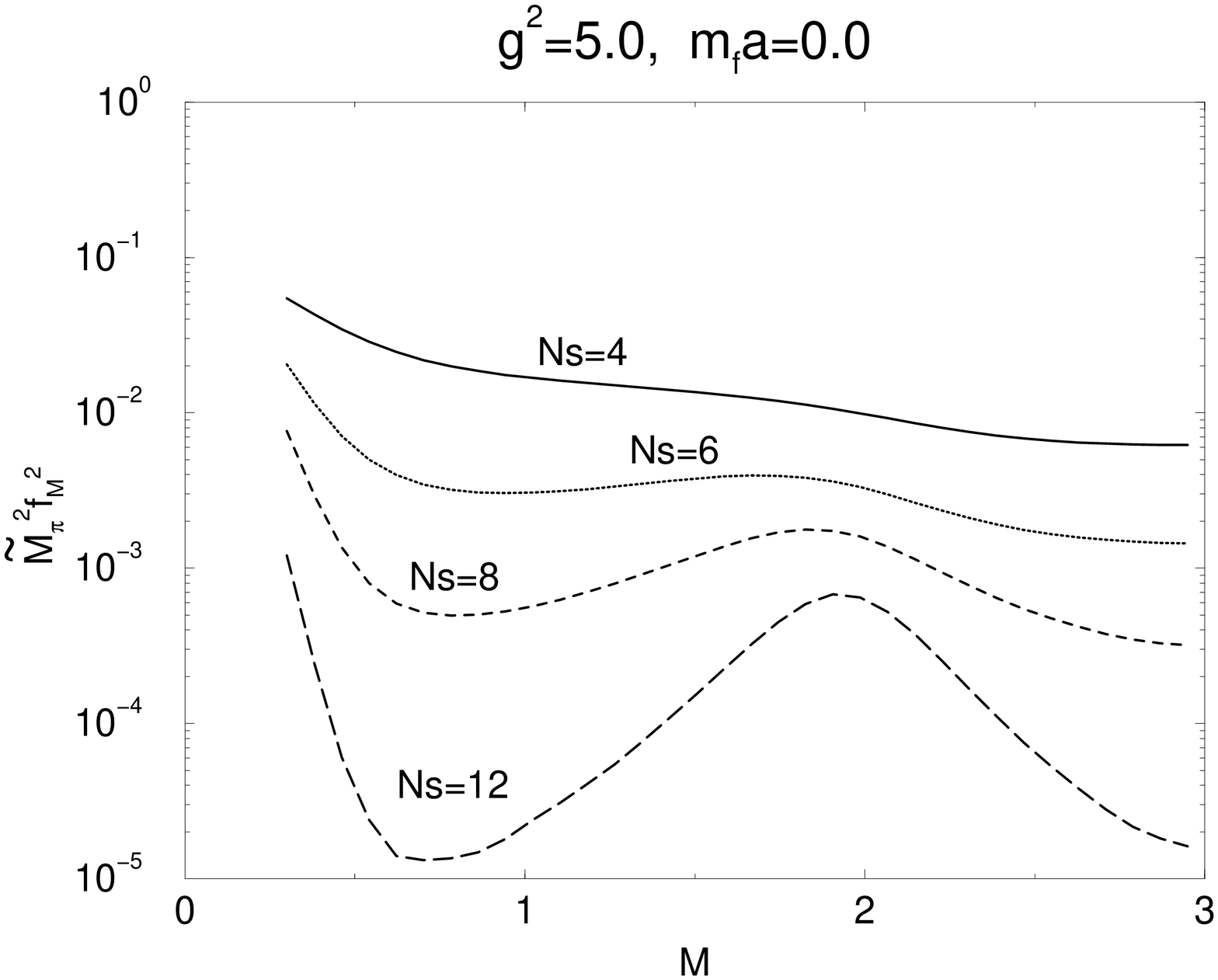}}
\caption{
$\widetilde{M}_\pi^2 f_M^2 $ as a function of $M$.
$g^2=1.0$ and $g^2=5.0$  for $N_s=2,4,6,8,12$.
}
\label{fig:Mpi2_M-g1.0}
\end{figure}

The systematic dependences of observables to the parameter $M$ are
shown in Figs.\ref{fig:pbp_M-g1.0} and \ref{fig:Mpi2_M-g1.0}.
In Fig.\ref{fig:pbp_M-g1.0}
we plot  $\langle\bar q q\rangle$ as a function of
$M$ for several $m_f$ with $g^2=1$ and $N_s=20$. 
$\langle\bar q q\rangle$ continuously gains in
magnitude with increasing $M$ around $M\sim 1$.
$\langle\bar q q\rangle$ for $m_f=0$ turns out to have
minimum magnitude around $M\sim 2$.

$\widetilde{M}_\pi^2$  also has the systematic dependence to $M$.
The ratio $\widetilde{M}_\pi^2/M_\pi^2$ is a smooth function of $M,N_s$ 
and $\widetilde{M}_\pi^2$ shows essential characteristics of 
the pion mass.
In Fig.\ref{fig:Mpi2_M-g1.0}
$\widetilde{M}_\pi^2 f_M^2 = d^2 V_{eff} / d \Pi^2$ is plotted 
against $M$.
Comparing results of strong coupling ($g^2=5.0$) with 
that of weak coupling ($g^2=1.0$) in this figure, 
we find that the depression of $\widetilde{M}_\pi^2 f_M^2$ near 
$M\sim1$ for weak coupling is not very manifest 
for strong coupling especially in small $N_s$.
In fact the minimum of $\widetilde{M}_\pi^2 f_M^2$ places 
at $M > 2$ for $N_s=4$.
This implies that in order to find the allowed region of $M$ 
for the chiral continuum limit
by observing the depression of pion mass, one needs larger $N_s$ for
strong coupling than that needed for weak coupling in DWQCD simulations.
Note that such allowed region of $M$ is unknown a priori in QCD by the
additive quantum corrections.

As we pointed out in previous section, 
pion mass does not vanish for all $M$ at $m_f=0$ for finite $N_s$ 
if $g^2 > 0$.
For  $N_s=4$, $\widetilde{M}_\pi^2$ at $m_f=0$ has its minimum at around 
$M=1.4$. 
For larger $N_s$, $\widetilde{M}_\pi^2(m_f=0)$ in the region $0<M<2$ tends to
be flat with smaller pion mass. 
From this figures we conclude that 
chiral symmetry is restored for large $N_s$
if $M$ is set in the region $0<M<2$. 
There is another region $2<M<4$, 
where pion would be massless particle in $N_s\to\infty$ limit.
This region corresponds to the region of massless ``pion'',
which is  made of the chiral modes at $p_\mu a = (\pi,0), (0,\pi)$.

\section{Continuum limit and disappearance of \Oa scaling violations}
\label{sec:Aerr}

\begin{figure}
\mbox{\epsfxsize=7.5cm \epsfbox{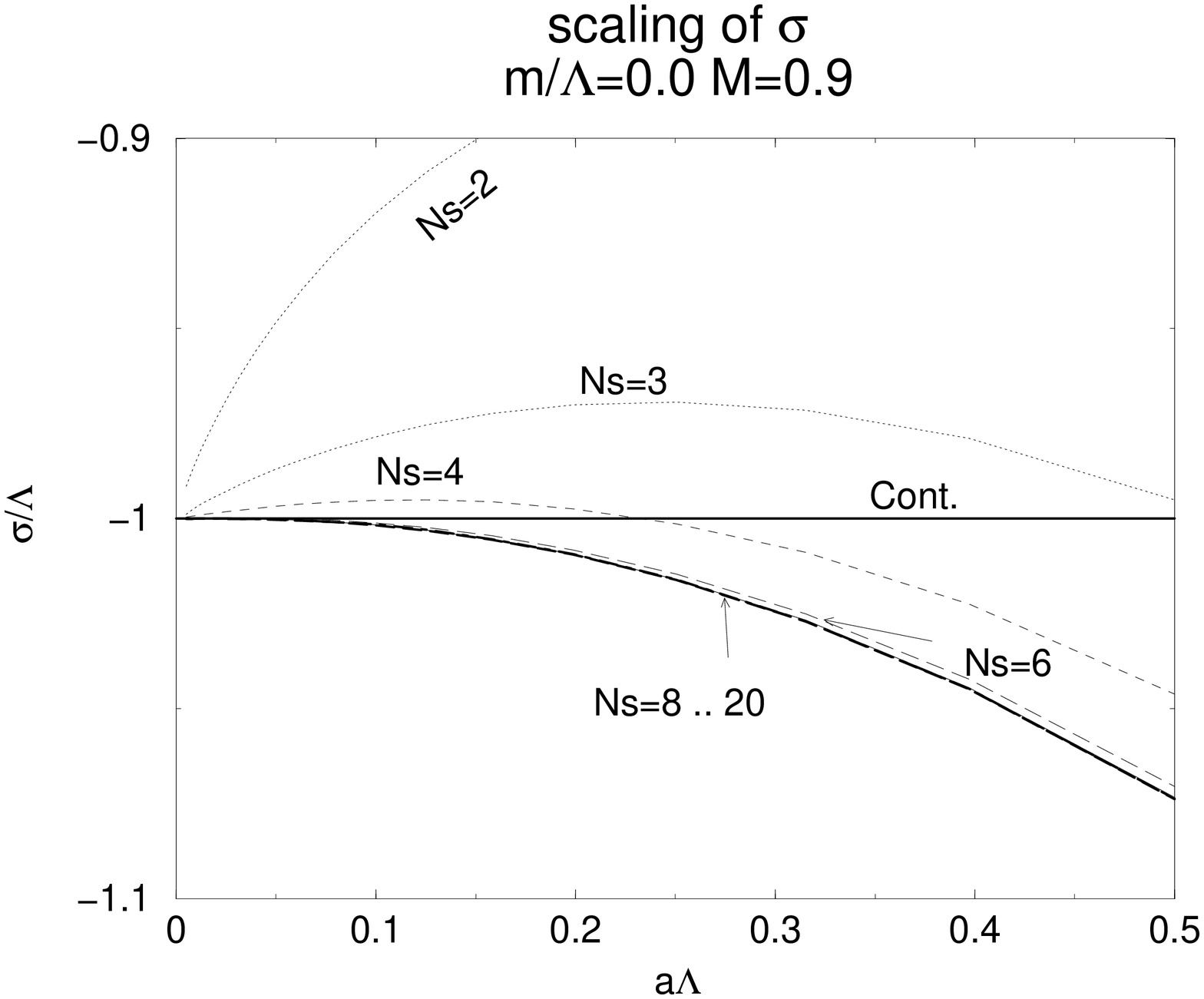}}
\mbox{\epsfxsize=7.5cm \epsfbox{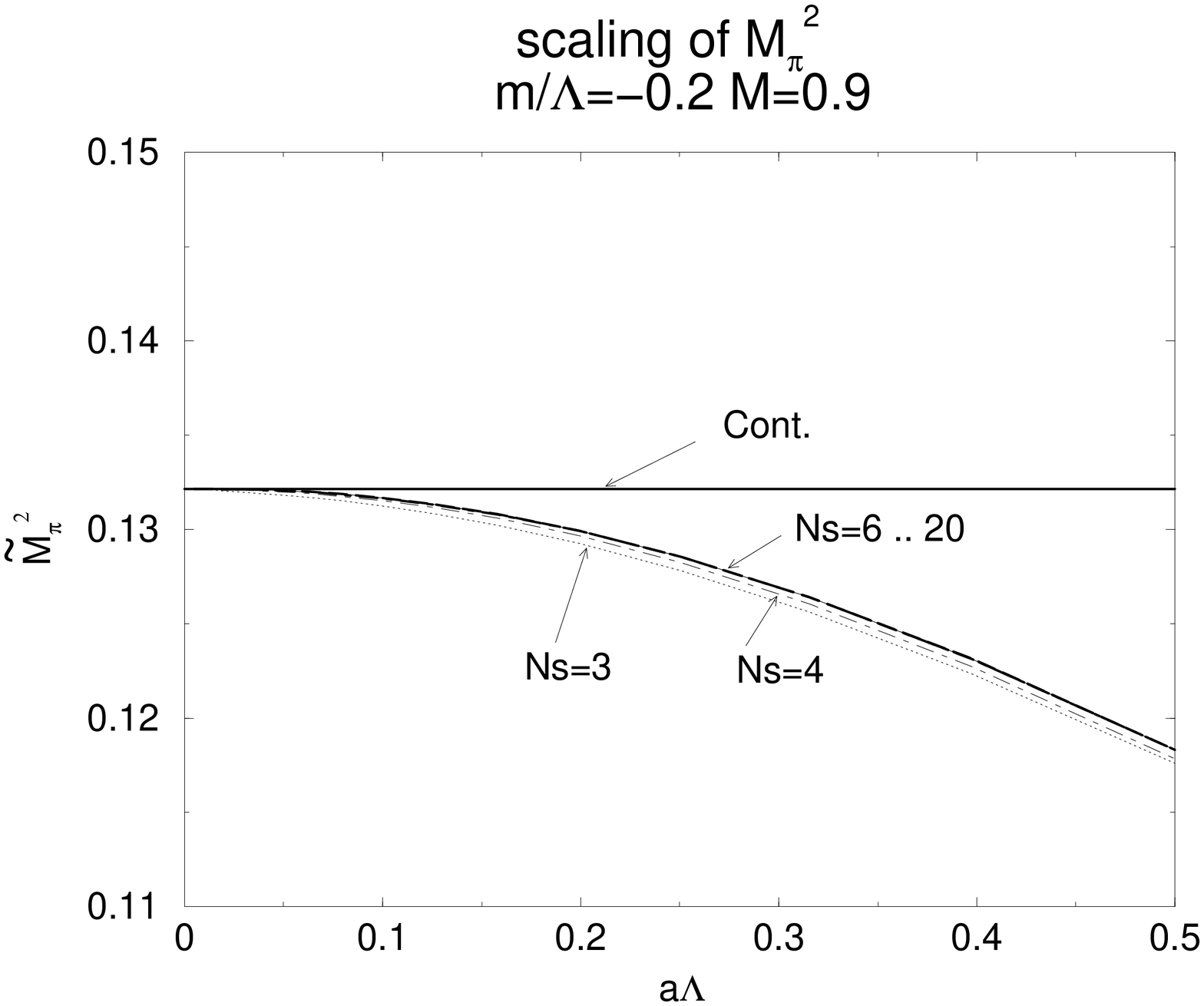}}
\caption{
$\sigma/\Lambda$ and $\widetilde{M}_\pi^2$ as a function of $a\Lambda$
for fixed $M$. 
The continuum limit is taken using the 
(Wilson like) scaling relations in
(\ref{eq:gsgtuning}-\ref{eq:tuning}).
}
\label{fig:Obs_a-Mfix}
\end{figure}

Toward the continuum limit $a\to0$, the lattice bare parameters 
are tuned according to the scaling relation 
(\ref{eq:gsgtuning}-\ref{eq:tuning}) for finite $N_s$.
We plot $\sigma$ and $M_\pi^2$ as a function of the lattice spacing
$a\Lambda$ in Fig. \ref{fig:Obs_a-Mfix} for several $N_s$ with 
fixed DW-mass, $M\sim 1$.

The lattice bare observables systematically depend on  $M$ 
as shown in previous section. 
From (\ref{eq:gpituning}) 
the lattice spacing (or the scale parameter of the theory) is
also a function of $M$,
\begin{equation}
\Lambda a(M,N_s) = \exp\left[ 
-{2\pi\over f_M^2}\left( {1\over2g_\pi^2} -\hat C_0(M,N_s)\right) \right]~.
\end{equation}
See Fig.\ref{fig:coef_c0}
for $(M,N_s)$ dependence of $\hat{C}_0$.
The $M$ dependencies of observables cancel with that of the lattice spacing,
and the correct continuum value are  reproduced at $a\to0$ limit.
(See Fig.\ref{fig:Obs_a-Mfix}.)

The interesting observation is the disappearance of $O(a)$ 
scaling violation for large $N_s$ limit. 
From theoretical point of view, the exact chiral symmetry 
in $N_s\to\infty$ limit is expected to prohibits
the dimension $(D+1)$ operators in the quantum correction, 
which cause $O(a)$ scaling violation.
The slope at $a\Lambda=0$  curve in Fig. \ref{fig:Obs_a-Mfix}
is finite at $N_s=2$ while
the curve at $N_s=20$ is flat near $a\Lambda=0$.
This shows that $O(a)$ scaling violation vanishes in large $N_s$ limit and
the scaling violation proportional to $a^2$ exists.

\begin{figure}
\mbox{\epsfxsize=7.5cm \epsfbox{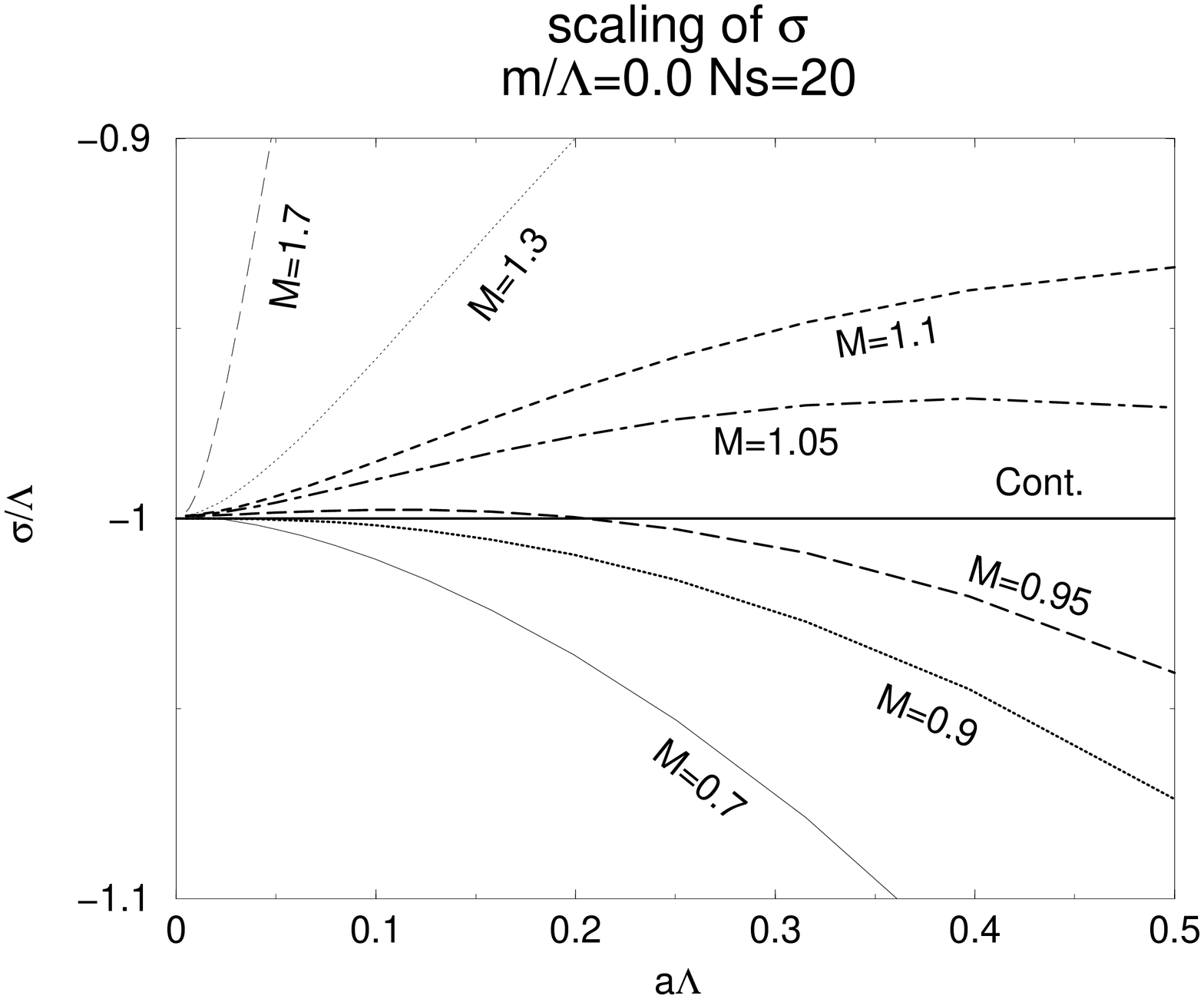}}
\mbox{\epsfxsize=7.5cm \epsfbox{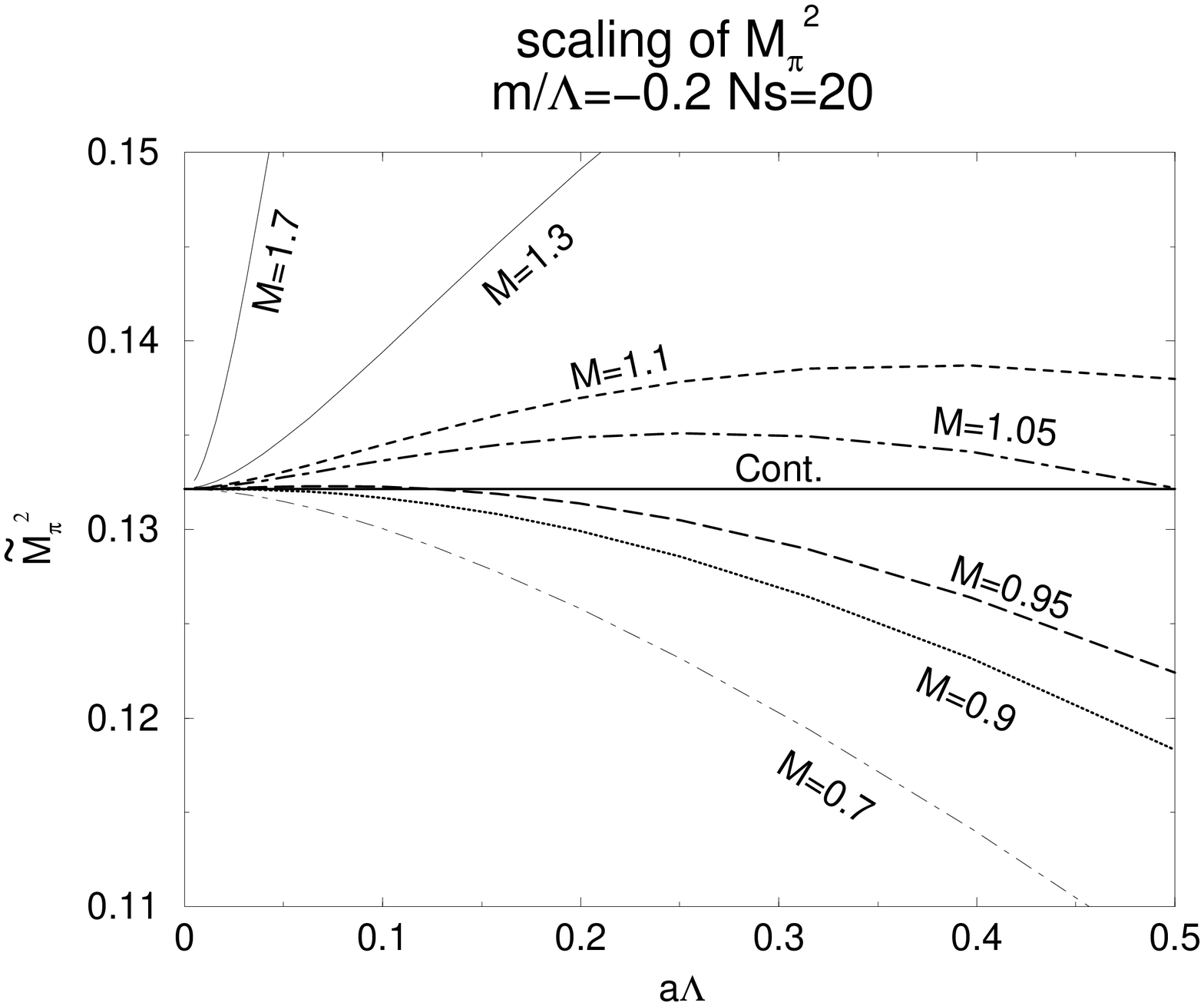}}
\caption{
$\sigma/\Lambda$ and $\widetilde{M}_\pi^2$ as a function of $a\Lambda$
at $Ns=20$. 
}
\label{fig:Obs_a-M0.9}
\end{figure}

Fig.\ref{fig:Obs_a-M0.9} shows that 
the remaining $O(a^2)$ error is small for $M\sim 1$ 
in large $N_s$, 
which is less than a few percentage for $a\Lambda < 0.5$ 
in this model.
The reason why this $O(a^2)$ scaling violation is small near $M=1$ 
could be understood by expanding 
the inverse integrand of the function ``$I_0(\sigma,\Pi)$''
for small $a$:
\begin{equation}
\frac{H^\prime}{F a^2} = p^2/f_M^2 + c p^4/f_M^4 a^2 + c' p^6/f_M^6 a^4 
+ \cdots
\end{equation}
Since $1/f_M$ is minimum  at $M=1$ so the $O(a^2)$ deviation from
the continuum formula is also minimum for $M=1$.
This feature might be similar for QCD simulation, except $M$ receives the
additive renormalization from the quantum fluctuation of the gauge field.

\begin{figure}
\mbox{\epsfxsize=7.5cm \epsfbox{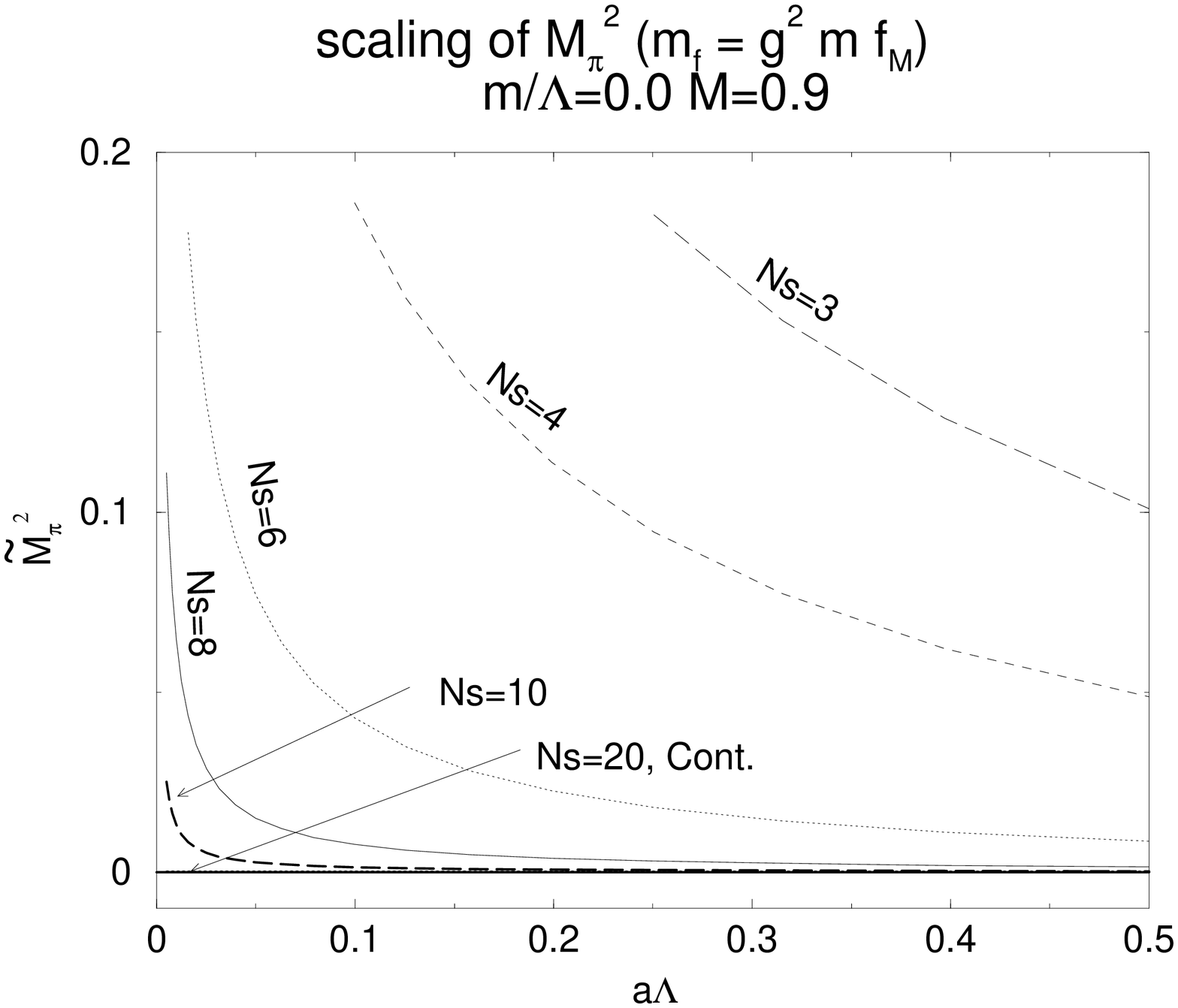}}
\mbox{\epsfxsize=7.5cm \epsfbox{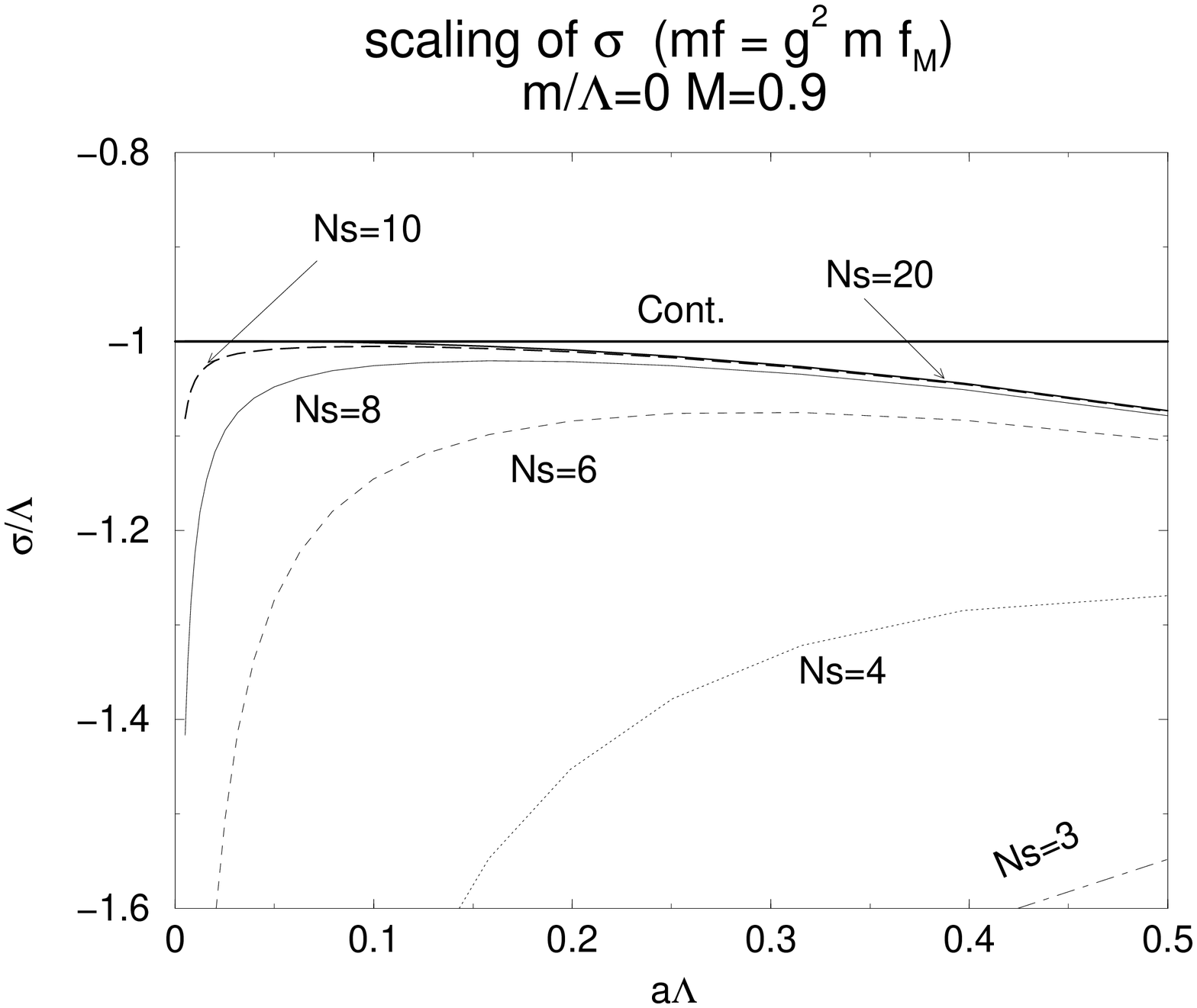}}
\caption{
$\sigma/\Lambda$ and $\widetilde{M}_\pi^2$ as a function of $a\Lambda$
at $M=0.9$. 
The DWF scaling relations 
in (\ref{eq:tuningDW1}-\ref{eq:tuningDW3}) 
are employed. 
}
\label{fig:Obs_aDWF-M0.9}
\end{figure}

The renormalization formula  employed above for finite $N_s$  is similar 
to that of WF. 
The bare mass needs to be fine tuned toward a certain critical value. 
One the other hand, we consider to apply 
the scaling relation for $N_s\to\infty$ (without fine tuning),
\begin{eqnarray}
&&\frac{1}{2 g_\sigma^2} - \hat{C_0} + C_2 
= \frac{f_M^2}{4 \pi} \ln \frac{1}{a^2 \Lambda^2} , 
\label{eq:tuningDW1}\\
&& g_\pi^2 = g_\sigma^2
\label{eq:tuningDW2}\\
&&\frac{1}{f_M} \frac{m_f}{g_\sigma^2} = m .
\label{eq:tuningDW3}
\end{eqnarray}
to {\it finite} $N_s$ lattice observables.
This is similar to what is done in the DWQCD simulation in a sense.
The result of this calculation for finite $N_s$
is shown in Fig.\ref{fig:Obs_aDWF-M0.9}.
For each $N_s$,  $\widetilde{M}_\pi^2$ is apt to go to 
the correct continuum value for large $a\Lambda$ but 
tends to diverge for smaller lattice spacing.
Such divergent behavior is never seen 
in current DWQCD simulations\cite{Simulations1,Simulations2} 
since $a\Lambda_{\rm QCD}$ is larger than 0.1.

\begin{figure}
\mbox{\epsfxsize=7.5cm \epsfbox{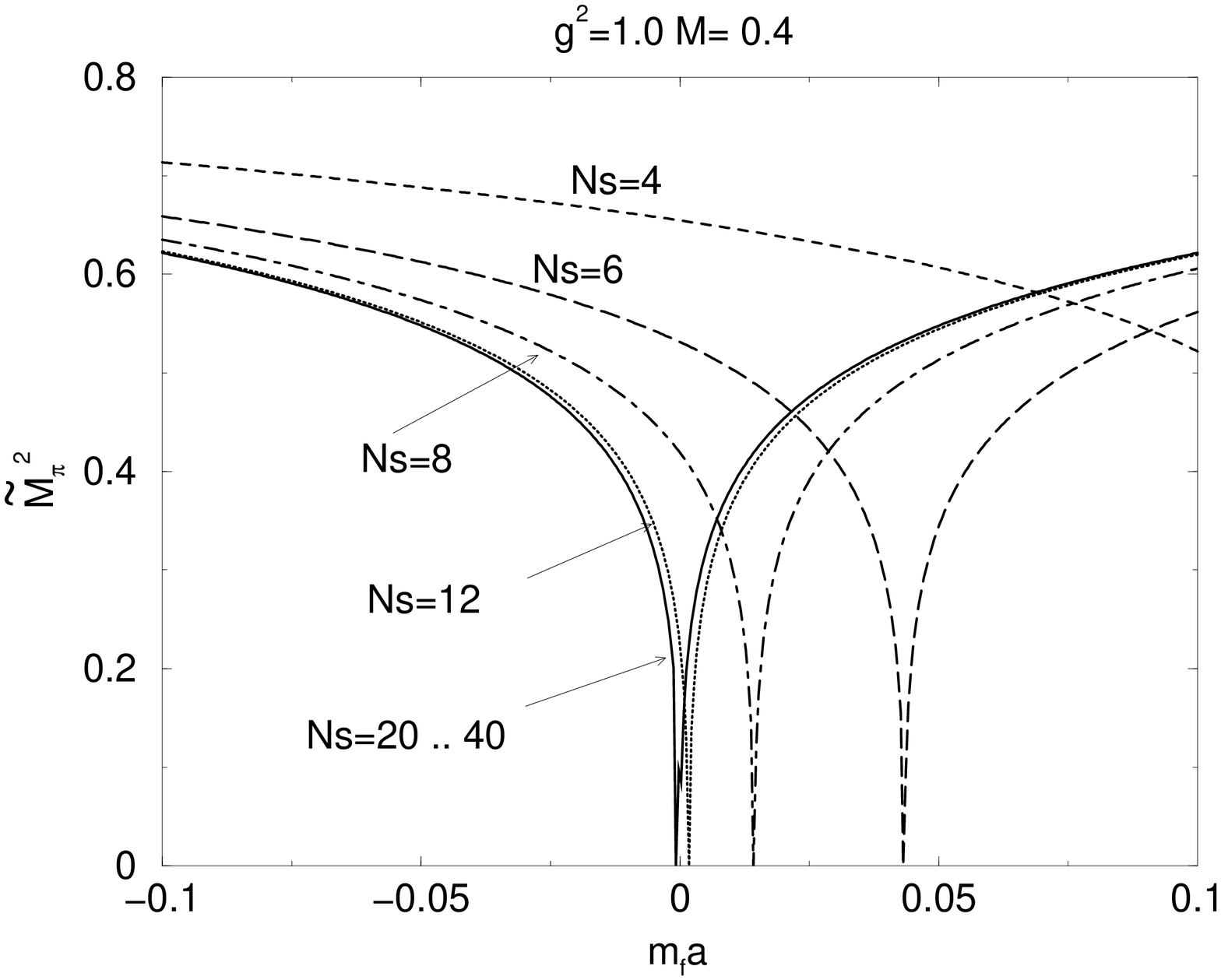}}
\mbox{\epsfxsize=7.5cm \epsfbox{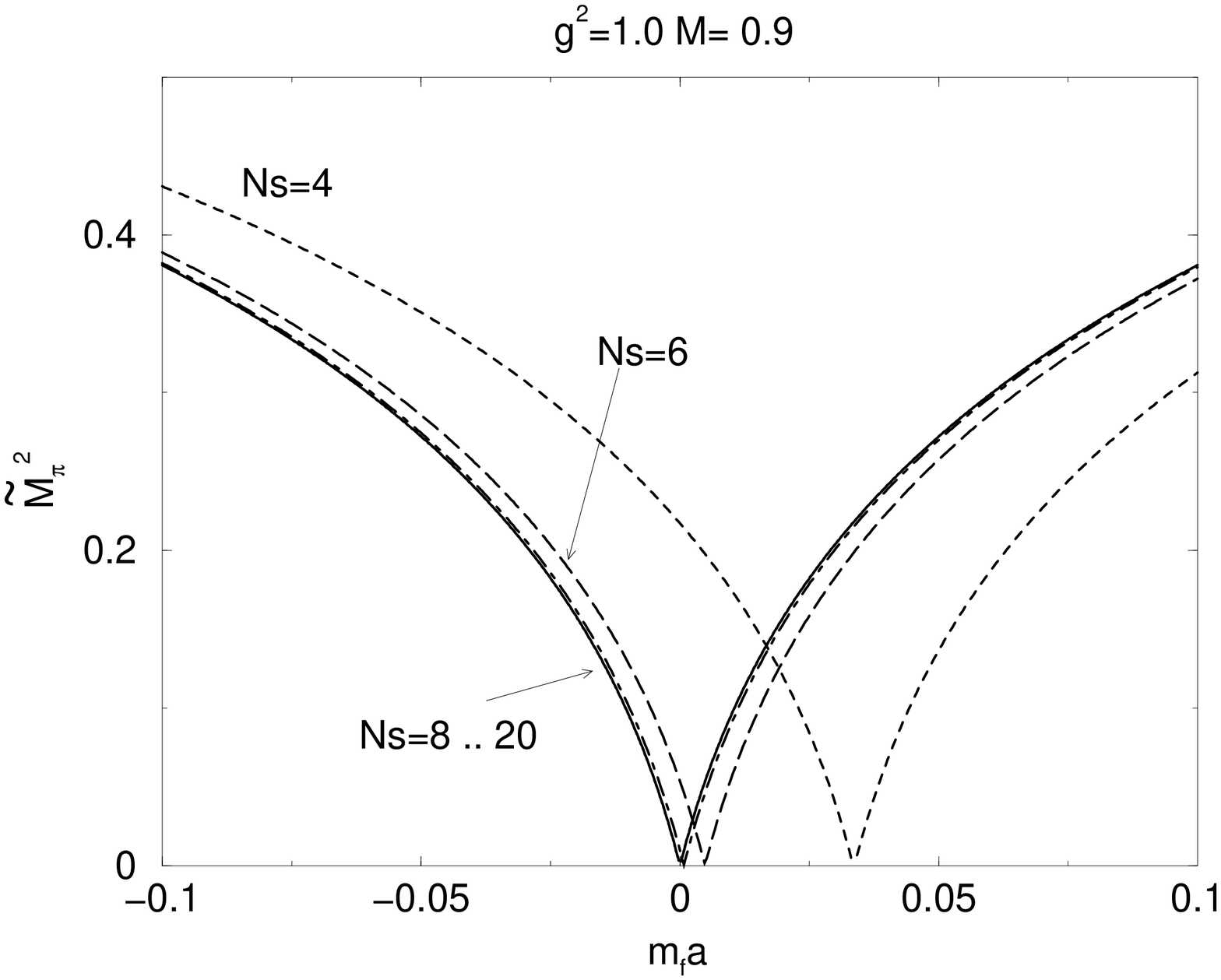}}
\caption{
$N_s$ dependence of  $\widetilde{M}_\pi^2$ as a function of $m_f$.
}
\label{fig:Mpi2_mf-Mfix}
\end{figure}

From (\ref{eq:tuning}) the renormalized mass is expressed as a
function of the bare quark mass $m_f$:
\begin{equation}
m a = {1\over f_M g^2} \left[ m_f a - (1-M)^{N_s} + g^2 C_1(M,N_s) \right] ~.
\end{equation}
For each $g^2$ (and $a\Lambda$), if $N_s$ fulfills a condition;  
\begin{equation}
N_s > N_s^{c}  \quad {\rm s.t.} \quad   |a m_f| 
\gg | -(1-M)^{N_s^{c}} + g^2 C_1(M,N_s^{c})|~,
\label{eq:NsCAnsatz} 
\end{equation}
the renormalized mass approximately becomes that in $N_s\to\infty$ case.
Such $N_s^{c}$ can exist for finite $m_f$ if $0<M<2$ 
because both $(1-M)^{N_s}$ and $C_1(M,N_s)$ go to zero for large $N_s$.
This leads that if $N_s$ is larger than $N_s^{c}$, physical
predictions from DWF is saturated as function of $N_s$ and could be 
regarded as the value of $N_s\to\infty$.
$N_s^{c}$ is a function of $g^2, M$ and $m_f$, and tends to be larger for
smaller $m_f$.
$N_s$ dependence of  $\widetilde{M}_\pi^2$ could be seen in  
Fig.\ref{fig:Mpi2_mf-Mfix}.
For large magnitude of $m_f$ the $N_s$ dependence is saturated up to
$N_s\sim 12$, while $\widetilde{M}_\pi^2$ varies as a function 
of $N_s$ near $m_f= 0$. 
From this figure for $(g^2,M)=(1,0.4)$, we can estimate 
$N_s^{c}$ is near $10$ for $|m_f| \sim 0.1$ and $N_s^{c} \sim 20$ for
$|m_f| \sim 0.02$. 
In this model $N_s^{c}$ goes to small value for $M\sim 1$. 
For $M=0.9$ the value of $\widetilde{M}_\pi^2$ for $N_s=12$ is 
nearly identical to that of $N_s=20$ for almost all the region of $m_f$ 
in Fig.\ref{fig:Mpi2_mf-Mfix}.

\section{Conclusions and Discussions}
\label{sec:conclusion}

We have investigated the two dimensional lattice GN model 
with DWF in large flavor($N$) limit,
as the toy model of lattice QCD with DWF.
By calculating the effective potential
we study the nonperturbative prospects of this model,
which are expected to be qualitatively similar to the DWQCD.

In infinite $N_s$ case, 
the effective potential has exact chiral symmetry 
even for finite lattice spacing.
The chiral phase boundary places exactly on $m_f=0$ line, 
which shows the fine tuning of the mass parameter,$m_f$, becomes needless.
The parity broken phase does not exist for all coupling constants.
Thus the model for $N_s=\infty$ has similar properties 
as the continuum theory especially for chiral symmetry,
by which the massless pion could be understand as  NG--boson 
accompanying  with the spontaneous breakdown of the symmetry.

In finite $N_s$ case, for which numerical simulations are carried out,
is practically important.
Chiral symmetry is explicitly broken by the finite $N_s$ effect,
that causes the parity broken phase with $(D+1)$ cusps near $g^2=0$ 
similar to Aoki phase of WF.
The restoration of chiral symmetry only occurs in the continuum limit
with fine tuning of $m_f$ to its critical point, which is on the phase
boundary of the parity broken phase.
By increasing $N_s$ for fixed $a$, 
the phase boundary exponentially approaches to $m_f=0$ plane 
in the parameter space and the parity broken phase vanishes 
in $N_s\to\infty$.  
If one take the limit $N_s\to\infty$  prior to
$a\to0$ limit, chiral symmetry is restored. 

Another interesting observation is the $N_s$ dependence of the 
critical  coupling, $g^2_c$. For $g^2 < g^2_c$,  the parity broken phase 
splits into $(D+1)$ regions, each of which corresponds to 
a chiral symmetric continuum limit.
The restoration of the chiral symmetry is also characterized by 
the fact that $g^2_c$ exponentially goes to large value 
with increasing $N_s$.

We also show ($M,N_s$) dependencies of lattice bare observables,
which could be some hints for DWQCD simulations.
When $N_s$ is finite and $g^2>0$, 
the pion mass at $m_f=0$ never goes to zero for all $M$.
If $M$ is set into the range $(0,2)$, 
the pion mass at $m_f=0$ exponentially suppressed with increasing $N_s$.
From the results of $m_f$ at $M_\pi=0$ (Fig.\ref{fig:mfM})
and $M_\pi$ at $m_f=0$ (Fig.\ref{fig:Mpi2_M-g1.0}),
we discussed that one needs to take larger $N_s$ for
strong coupling than that needed for weak coupling in DWQCD simulations,
in order to search the allowed region of $M$ 
for the chiral continuum limit
by observing the depression of pion mass, 
By observing the $N_s$ dependences of the observables we discuss about
the criterion, $N_s > N_s^{c}$, for which the physical observables 
can be considered as approximate values for $N_s\to\infty$.
$N_s^{c}$ is a function of $g^2, M$ and $m_f$.

The observables depend on the value of $M$ even for $N_s\to\infty$.
This dependence cancels by the renormalization and 
the correct continuum theory is obtained.
The disappearance  of $O(a)$ scaling violation for large $N_s$ 
in the continuum limit suggests the probability
of obtaining the reliable physical predictions 
for smaller lattice spacing than that in WF. 

Since the GN model in large $N$ limit
neglects the quantum fluctuations, and omits the gauge fields,
we cannot insist that the behavior of lattice QCD with DWF
is exactly same as the results in this paper.
For example, the DW-mass, $M$, is shifted into $\widetilde{M}=M+Const.$
due to the back reaction of the gauge fields.
However there was many similarities between 
GN model and QCD for Wilson action\cite{aoki,umemura,kaneda},
we expect that 
the results shown in this paper will provide  
instructive and systematic informations
about {\it nonperturbative} effects of lattice QCD with DWF.

\section*{Acknowledgements}
We thank S. Aoki, Y. Kuramashi, Y. Taniguchi and A. Ukawa for 
illuminating discussions and continuous encouragements. 
Numerical calculations for the present work have been carried out
at Center for Computational Physics at University of Tsukuba. 
This work is supported in part by the Grants-in-Aid 
for Scientific Research from the Ministry of Education, 
Science and Culture (Nos. 2375, 6769).
The authors are supported by Japan Society for Promotion of Science.

%

%
%
%
\newpage
%
%
%
%
%

\vspace*{1cm}

\vspace*{1cm}

\newpage


\vspace*{2cm}

\vspace*{1cm}

\vspace*{2cm}

\vspace*{1cm}

\vspace*{2cm}

\end{document}